\documentclass[12pt,preprint]{aastex} 

\usepackage{graphicx}
\usepackage[twocolumn]{emulateapj5}

\newcommand{\halpha}{H$\alpha$}
\newcommand{\hbeta}{H$\beta$}
\newcommand{\hgamma}{H$\gamma$}

\newcommand{\sameauthor}{\underbar{\qquad\qquad}.}

\shortauthors{SCHMIDT ET AL.}
\shorttitle{LOW ACCRETION-RATE BINARIES}

\received{}
\revised{}
\submitted{Accepted to the Astrophysical Journal, Part 1}

\begin{document}

\tolerance 10000

\title{New Low Accretion-Rate Magnetic Binary Systems and their Significance
for the Evolution of Cataclysmic Variables\altaffilmark{1,2}}

\altaffiltext{1}{A portion of the results presented here were obtained with
the MMT Observatory, a facility operated jointly by The University of Arizona
and the Smithsonian Institution.}
\altaffiltext{2}{Based in part on observations with
the Apache Point Observatory 3.5~m telescope and the Sloan Digital Sky Survey,
which are owned and operated by the Astrophysical Research Consortium (ARC).}
\author{
Gary D. Schmidt\altaffilmark{3},
Paula Szkody\altaffilmark{4},
Karen M. Vanlandingham\altaffilmark{3},
Scott F. Anderson\altaffilmark{4},
J. C. Barentine\altaffilmark{5},
Howard J. Brewington\altaffilmark{5},
Patrick B. Hall\altaffilmark{6},
Michael Harvanek\altaffilmark{5},
S. J. Kleinman\altaffilmark{5},
Jurek Krzesinski\altaffilmark{5,6}
Dan Long\altaffilmark{5},
Bruce Margon\altaffilmark{8},
Eric H. Neilsen, Jr.\altaffilmark{9},
Peter R. Newman\altaffilmark{5},
Atsuko Nitta\altaffilmark{5},
Donald P. Schneider\altaffilmark{10},
\and
Stephanie A. Snedden\altaffilmark{5}
}

\vskip 10pt

\altaffiltext{3}{Steward Observatory, The University of Arizona, Tucson, AZ
85721.} \email{gschmidt@as.arizona.edu}
\altaffiltext{4}{Department of Astronomy, University of Washington, Box 351580,
Seattle, WA 98195-1580.}
\altaffiltext{5}{Apache Point Observatory, P.O. Box 59, Sunspot, NM 88349}
\altaffiltext{6}{Department of Physics \& Astronomy, York University, 4700
Keele St., Toronto, ON, M3J 1P3, Canada.}
\altaffiltext{7}{Mt. Suhora Observatory, Cracow Pedagogical University, ul.
Podchorazych 2, 30-084 Cracow, Poland.}
\altaffiltext{8}{Space Telescope Science Institute, 3700 San Martin Drive,
Baltimore, MD 21218.}
\altaffiltext{9}{Fermi National Accelerator Laboratory, P.O. Box 500, Batavia,
IL 60510.}
\altaffiltext{10}{Pennsylvania State University, Department of Physics \&
Astronomy, 525 Davey Lab., University Park, PA 16802.}

\begin{abstract}

Discoveries of two new white dwarf plus M star binaries with striking optical
cyclotron emission features from the Sloan Digital Sky Survey (SDSS) brings to
six the total number of X-ray faint, magnetic accretion binaries that accrete
at rates $\dot M \lesssim 10^{-13}~M_\sun$ yr$^{-1}$, or $<$1\% of the values
normally encountered in cataclysmic variables. This fact, coupled with donor
stars that underfill their Roche lobes and very cool white dwarfs, brand the
binaries as post common-envelope systems whose orbits have not yet decayed to
the point of Roche-lobe contact.  They are pre-magnetic CVs, or pre-Polars.
The systems exhibit spin/orbit synchronism and apparently accrete by efficient
capture of the stellar wind from the secondary star, a process that has been
dubbed a ``magnetic siphon''.  Because of this, period evolution of the
binaries will occur solely by gravitational radiation, which is very slow for
periods $>$3~hr.  Optical surveys for the cyclotron harmonics appear to be the
only means of discovery, so the space density of pre-Polars could rival that of
Polars, and the binaries provide an important channel of progenitors (in
addition to the asynchronous Intermediate Polars).  Both physical and SDSS
observational selection effects are identified that may help to explain the
clumping of all six systems in a narrow range of magnetic field strength around
60~MG.

\end{abstract}


\section{Introduction}

The physics of radial accretion onto the surface of a white dwarf divides
naturally into regimes defined by the specific accretion rate and local
magnetic field strength (e.g., Lamb \& Masters 1979; Wickramasinghe \&
Ferrario 2000 and references therein).  For comparatively dense flows and weak
fields, $\dot m \gtrsim 1$ g cm$^{-2}$ s$^{-1}$, $B \lesssim 50$~MG, a strong
shock radiates primarily by the bremsstrahlung mechanism (emissivity
$\epsilon_{br} \propto N_e^2$) at a temperature common to both ions and
electrons of $kT\sim10-30$~keV.  The highest accretion rates, $\dot m \gtrsim
100$ g cm$^{-2}$ s$^{-1}$, are still cooled largely by hard X-rays, but the ram
pressure of the stream can depress the shock below the level of the
surrounding photosphere, so that the emergent reprocessed radiation appears
predominately in the soft X-rays/EUV. More rarefied streams and/or stronger
magnetic fields favor cyclotron emission ($\epsilon_{cyc} \propto N_e B^2$),
which results in a reduced shock height and two-fluid plasma with the
electrons being substantially cooler than the ions (Fischer \& Beuermann
2001).  At the typical total mass accretion rate of a magnetic cataclysmic
variable (CV; $\dot M \sim 10^{-11}-10^{-10}~M_\sun$~yr$^{-1}$; e.g., Warner
1995), the conditions in the accretion region(s) are generally spanned by the
above parameter range, and strong X-ray and polarized optical cyclotron
emission are observed.  In view of this fact, it is no coincidence that the
current catalog of magnetic CVs (Ritter \& Kolb 2003) is dominated by
discoveries from orbiting X-ray observatories (especially {\em ROSAT}).

The recent initiation of large-area, deep optical surveys -- generally targeted
at the extragalactic universe -- has uncovered a sample of six accreting
magnetic binaries that display remarkably prominent, detached, and
circularly-polarized cyclotron emission harmonics atop the stellar continua
(Reimers et al. 1999; Reimers \& Hagen 2000; Szkody et al. 2003, hereafter
S03).  The almost complete lack of X-ray emission from these systems (Szkody et
al. 2004; hereafter S04) explains how they eluded the earlier all-sky X-ray
surveys and, together with the optical spectral appearance, implies an
extremely low specific accretion rate, $\dot m \sim 0.001-0.01$ g cm$^{-2}$
s$^{-1}$.  Here, theory indicates that no shock forms (Kuijpers \& Pringle
1982). This ``bombardment'' regime is a non-hydrodynamic solution that
describes conditions in which the heated electrons cool on a time scale short
compared with the stopping time of the protons (e.g., Woelk \& Beuermann 1996
and references therein).  In the case of a strongly magnetic white dwarf, the
cooling occurs primarily by cyclotron emission at a low atmospheric column
depth and at a temperature well below that of a radial shock ($kT \sim
1$~keV).  Thus, virtually all of the accretion energy emerges in a few isolated
emission features at low harmonic number.  Though the very low-$\dot m$ regime
was explored theoretically to explain observations of intense cyclotron lines
in known AM Her systems during temporary episodes of low mass transfer (total
accretion rate $\dot M \sim10^{-12}~M_\sun$~yr$^{-1}$), it appears that the
newly-discovered objects represent a distinct, possibly evolutionary, binary
phase with even more extreme and persistent characteristics.  Indeed, Schwope
et al. (2002) have coined the term Low Accretion Rate Polar (LARP) to highlight
their potential importance.

In this paper we report the discovery of two additional low-$\dot m$ magnetic
binaries from recent Sloan Digital Sky Survey (SDSS) observations.  Followup
photometric, spectroscopic, and spectropolarimetric observations of these and
previously-discovered examples are analyzed to gain insight into the currently
small but intriguing class of objects. From this evidence we explore ideas as
to their evolutionary state, space density, and relevance to both magnetic and
nonmagnetic CVs.  Preliminary discussion of these results was presented by
Schmidt (2004), and some of the conclusions were reached by Webbink \& Wickramasinghe (2004) from a more theoretical perspective.

\section{Observational Data}

The SDSS has proven to be an extremely valuable, if not optimized, tool for
identifying various classes of stellar systems.  More than 120 new CVs have
been cataloged thus far (Szkody et al. 2005 and references therein), based on
spectroscopy of targets selected by algorithms that are sensitive to regions of
the color-color planes occupied by hot stars, QSOs, white dwarfs and white
dwarf-M star pairs.  Indeed, because of the presence of intense, narrow,
cyclotron harmonics, three of the four SDSS low-$\dot m$ magnetic accretion
binaries were targeted as candidate QSOs.  The exception is
SDSS~J2048+0050\footnote{For brevity, we will refer to objects in the text and
figures by the designation SDSS~J{\em hhmm}$\pm${\em ddmm}.}, which exhibits
comparatively weak features and was targeted as part of the Sloan Extension for
Galactic Understanding and Evolution (SEGUE; e.g., Beers et al. 2004).  As an
illustration, we compare the locations of the four SDSS low-$\dot m$ systems to
other classes of objects in the SDSS $u-g$, $g-r$ plane in Figure 1.

The SDSS spectroscopic data are obtained with twin dual-beam spectrographs
covering the regions $3900-6200$~\AA\ and $5800-9200$~\AA\ at a resolving power
of $\lambda/d\lambda\sim1800$. The wide, continuous spectral coverage provides
broad sensitivity to cyclotron emission features, which may be prominent at
only one or two harmonics of the cyclotron fundamental frequency
$\omega_c=eB/m_ec$. Details of the SDSS photometric and spectroscopic hardware,
as well as the data reduction procedures and targeting strategy, can be found
in Fukugita et al. (1996), Gunn et al. (1998), Hogg et al. (2001), Lupton et
al. (1999, 2001), York et al. (2000), Pier et al. (2003), Smith et al. (2002),
and Ivezic et al. (2004).

Followup observations of known magnetic systems and confirmation of new
candidates were made through optical spectroscopy and circular
spectropolarimetry with the instrument SPOL (Schmidt et al. 1992) attached to
the 6.5 m MMT atop Mt. Hopkins and the Steward Observatory 2.3 m Bok telescope
on Kitt Peak.  Polarimetric data are essential because of the $\sim$141 million
objects imaged thus far by the SDSS, it is inevitable that there will be chance
superpositions of QSOs and late-type Galactic stars whose total flux spectra
mimic the binaries discussed here.  As an example, SDSS~J092853.51+570735.4
shows only marginal evidence for duplicity in the SDSS imaging, yet in the
survey spectrum displayed in the bottom panel of Figure 2, exhibits a single
emission feature at 4140\AA\ and the molecular band structure of an M dwarf.
In this respect it resembles the other two objects depicted in Figure 2.
However, SDSS~J0928+5707 shows no circular polarization to an upper limit of
$v=V/I=0.4\%$, and as also shown in the figure, subtraction of the spectrum of
an M3 V spectrum reveals the blue continuum and emission lines of a $z=1.67$
QSO.  SDSS~J164209.57+213352.8, with $z=1.43$, and probably
SDSS~J155839.64+082720.1 and SDSS~J105322.24+535510.1, are other cases that
have been encountered thus far.

In the configuration used, the 1200$\times$800 pixel SITe CCD used in the
spectropolarimeter provides a coverage $\sim$$\lambda\lambda4200-8400$ and
resolution $\sim$15~\AA.  The thinned, back-illuminated, and anti-reflection
coated device features broad sensitivity with quantum efficiency reaching 92\%
and a read noise of 2.3$e^-$.  Circular spectropolarimetry is obtained from
4-exposure sequences in different orientations of a quarter-waveplate, with a
Wollaston prism located in the collimated spectrograph beam splitting the
light into the complementary senses of polarization required for measurement
of a Stokes parameter.  Total flux spectra are derived from the sum of the
polarized spectra, with calibrations based on observations of spectral flux
standard stars made in the identical instrument configuration on the same
night, and terrestrial absorption features removed through observations of
proximate early-type stars.   Data reduction is carried out using custom
{\it IRAF\/} scripts\footnote{{\it IRAF\/} is distributed by the National
Optical Astronomy Observatories, which are operated by the Association of
Universities for Research in Astronomy, Inc., under cooperative agreement with
the National Science Foundation.}.

Two runs of CCD photometry were made on SDSS~J0837+3830 on 2004 Nov. 30 and
Dec. 1 in an effort to derive a photometric period.  These observations were
made in unfiltered light at the MDM 2.4~m telescope on Kitt Peak with the
Templeton CCD (a 1024$\times$1024 24$\mu$m pixel, thinned, back-illuminated
SITe CCD). Seeing was poor, averaging $\sim$4\arcsec\ FWHM on the first night
and 6\arcsec\ on the second, and the latter run was cut short by clouds.
Integration times were 300~s throughout.  The data frames were reduced using
standard {\it IRAF\/} routines and field stars were used to perform the
relative photometry.

Finally, time-resolved spectra, using 15 min integrations over the course of
4.3 hrs, were acquired for SDSS~J2048+0050 on 2004 October 4 using the Apache
Point Observatory (APO) 3.5 m telescope. The Double Imaging Spectrograph was
used with the 1.5\arcsec\ slit in high-resolution mode, resulting in a
resolution of $\sim$3\AA\ and wavelength coverage of $4000-5200$\AA\ in the
blue and $6000-7600$\AA\ in the red. Reduction of the spectra to wavelength and
flux was accomplished with standard {\it IRAF\/} routines using lamps and
standard star frames obtained during the same night.  Radial velocity and flux
measurements of the \halpha\ and \hbeta\ lines were made with the ``e'' routine
(which determines the line centroid) in the {\it IRAF\/} SPLOT package.  A log
of all of the above observations is provided as Table 1.

\section{Two New Low Accretion-Rate Magnetic Binaries}

Two new binaries with prominent, isolated emission features have been
identified from SDSS spectroscopy through the end of 2004.  In coordinate and
Plate-MJD-Fiber notation, they are: SDSS~J083751.00+383012.5 = 828-52317-049,
from Data Release 3 (DR3; Abazajian et al. 2005) and reported by Szkody et al.
(2005), and SDSS~J204827.91+005008.9 = 1909-53242-601.  The latter object will
be included in a future SDSS data release. Photometry is provided as SDSS {\it
psf\/} magnitudes in Table 2.

In the SDSS spectra shown in Figure 2, distinguishing humps between 4000\AA\
and 5000\AA\ are evident in both objects, but basic differences between these
spectra and those of previous low-$\dot m$ magnetic systems (Reimers \& Hagen
1999; Reimers et al. 2000; S03) highlight the wide variety of the class: The
hump in SDSS~J0837+3830 dominates the underlying stellar continuum with a
broad and rather flat-topped profile (FWHM $\gtrsim$ 1000\AA), and it is
accompanied by prominent narrow Balmer emission lines up to at least \hgamma.
Though the lines are likely more indicative of the conditions on the M-star
surface than of any accretion stream, all of these characteristics are
consistent with an accretion rate that may be slightly higher than has been
estimated for the previous systems ($\dot M \sim 10^{-13}~M_\sun$~yr$^{-1}$).
In contrast, the single hump in SDSS~J2048+0050 is weak compared with the cool
companion, and even though a narrow \halpha\ line is evident, \hbeta\ is
barely detected.  If these features are characteristic of the binary over long
time scales, the accretion rate would seem to be extremely low.

\subsection{SDSS~J0837+3830}

Each of the two runs of CCD photometry on SDSS~J0837+3830 shows a clear
brightness variation on a timescale of a few hours.  The two nights were
therefore combined and power spectrum analysis performed.  The result displays
two peaks of equal significance at periods of 3.18 and 3.65 hr, i.e., a
difference of one cycle between the two nights.  Least-squares fits to a sine
wave yield a semi-amplitude of 0.06~mag in either case, and inspection
suggests that the period with the slightly higher total $\chi^2$, 3.18 hr,
might actually be preferred because of fewer discrepant data points. This
choice is depicted in Figure 3.  Because the CCD was used unfiltered, it is
not possible to categorically assign the photometric variation to a changing
view of the magnetic field in the cyclotron emission region (the white dwarf
spin period) as opposed to the varying aspect of a tidally-locked secondary
star (the orbital period).  However, where information is available for the
previously-studied low-$\dot m$ systems listed in Table 3, indications are that
spin-orbit synchronism has been established\footnote{See also \S4}, and we
presume that this is the case also in SDSS~J0837+3830.  Unfortunately, the
interval between the photometric observations and our spectroscopy described
below is too long to place the latter data on a phase scale.

All four epochs of followup spectroscopy of SDSS~J0837+3830 yielded spectra
that mimic the survey data shown in Figure 2: a single dominant cyclotron hump
in the blue, narrow Balmer emission lines, and the continuum of a cool
companion star with $R\sim19.2$. Indeed, the observational record for this
object is as extensive as for the previously-discovered low-$\dot m$ magnetic
systems, and shows no evidence for high accretion states over a period of 3
yr.  The longest run with the spectropolarimeter spanned slightly more than 1.2
hr, or $\sim$40\% of the photometric period, while the object was setting on
the evening of 2004 May 13.  As depicted in Figure 4, each of the five 800~s
sequences displayed circular polarization of nearly $-$20\% in the 4500\AA\
hump, and a second negatively-polarized feature is present near 6200\AA. For a
low plasma temperature, these can be readily assigned to harmonics
$n=\omega/\omega_c=4$ and 3, respectively, in a field of 65~MG. However, the
shape of the $n=3$ circular polarization harmonic is distorted into a peaked
profile, and significant {\it positive\/} circular polarization, $v=2-4$\%, is
measured between the harmonics. This indicates that the oppositely-polarized
magnetic pole is also accreting, and that it may have a slightly different
field strength as well as a lower accretion rate. Variations through the
sequence are confined to a reduction in the emission-line strength and possible
fading of the $n=4$ harmonic.  The constancy in radial velocity of the
\halpha\ emission line to $\le$50~km~s$^{-1}$ and the small amplitudes of
polarimetric and photometric variability suggest that this binary may be
oriented at a low inclination.

\subsection{SDSS~J2048+0050}

The APO spectra of SDSS~J2048+0050 appear very similar to the SDSS spectrum
shown in Figure 2, which was obtained about 1.5 months earlier. The blue
spectra cover the cyclotron hump near 4550\AA\ and \hbeta, while the red
spectra follow \halpha\ and the TiO bands.  The Balmer lines are very narrow,
ruling out an origin in an infalling stream, and \ion{He}{2} $\lambda$4686 is
not present, indicating the lack of a hard irradiating continuum.  The
time-resolved spectra show a changing amplitude of the cyclotron hump and
strength of the Balmer emission, with the \hbeta\ line disappearing completely
in several spectra. Least squares fitting of the velocity measurements to a
sinusoid yields a period of 4.2 hr (250 min) for \halpha\ with a $K$
semi-amplitude of $153\pm5$ km s$^{-1}$ and rms deviation of 14 km s$^{-1}$
around the fit.  \hbeta\ was more difficult to measure and had fewer samples
due to its disappearance at some phases, so the period was fixed at 250 min to
derive a best fit solution with $K = 124\pm7$ km s$^{-1}$ and an rms of 18 km
s$^{-1}$. Again, our knowledge of the period is insufficient to link to other
datasets.  The fits are shown along with the data points and \halpha\ line
fluxes in Figure 5.  A least-squares sine fit to the fluxes indicates that the
peak flux is offset by $\sim$0.14 phase from the negative-to-positive velocity
zero-crossing that defines $\varphi=0$. The implications of these observations
are that the lines originate predominately from the inner hemisphere of the
secondary star, similar to the conclusion reached for WX LMi (HS~1023+3900) by
Schwope et al. (2002), but the emission is not symmetric around the line of
centers.  If the enhanced emission is caused by irradiation of the secondary by
a hot area near the accreting pole of the white dwarf, then it is likely that
the pole is not directly opposite the secondary.  Among the spin-orbit
synchronized magnetic CVs (Polars) it is common to find the principal magnetic
pole advanced somewhat prograde of the secondary in its orbit (Cropper 1988).

Spectropolarimetry of SDSS~J2048+0050 has thus far been limited to slightly
more than 1~hr obtained through cirrus on the 2.3~m telescope.  The data were
acquired in five sequences, and very little variation can be detected through
the series.  The combined flux spectrum, shown in Figure 6, again closely
resembles the SDSS spectrum in Figure 2 both in absolute brightness and in the
presence of a single hump centered near 4550\AA. \halpha\ is weak and
\hbeta\ is absent altogether at this combination of S/N and resolution.  The
net circular polarization in the 4550\AA\ feature is extremely high at
$-38$\%.  This, coupled with the fact that the peak flux of the feature is
elevated only $\sim$40\% above the stellar continuum, implies that the
cyclotron light in this harmonic must be essentially 100\% circularly
polarized!  The polarization also dips slightly below zero around 6200\AA,
suggesting the presence of a second, weaker cyclotron harmonic akin to that in
SDSS~J0837+3830. If real (see \S5), the harmonics have the same numerical
assignment as in that system and the magnetic field strength is very similar at
$\sim$62~MG.  No evidence exists for emission from the opposite magnetic pole,
at least during the phase interval recorded by the spectropolarimetry.

SDSS~J2048+0050 may be associated with a weak X-ray source in the {\it ROSAT\/}
All Sky Survey (RASS) at the level of 0.014 count s$^{-1}$, although there are
other optical objects within the error circle.  Using the crude conversion of 1
count s$^{-1}$ = $7\times10^{-12}$ erg cm$^{-2}$ s$^{-1}$ and a distance of
260~pc (derived below), a $0.1-2.4$~keV X-ray luminosity of
$L_X\sim8\times10^{29}$~erg s$^{-1}$ is found.  This would be approximately an
order of magnitude larger than the measurements of $L_X$ for SDSS~J1553+5516
and SDSS~J1324+0320 by S04, which were interpreted to arise from coronal
emission on the secondaries.  The result for SDSS~J2048+0050 is extremely
uncertain and begs for a deeper observation.

\section{New Results on Known Low-$\dot m$ Systems}

\subsection{Orbital Variations in SDSS~J1324+0320}

The cyclotron emission harmonics in SDSS~J1324+0320 display the greatest
contrast relative to the underlying continuum of any of the low-$\dot m$
systems discovered thus far.  The SDSS spectrum was presented by S03 together
with a single observation of confirming spectropolarimetry, and a magnetic
field strength of 63~MG was estimated from the identification of the humps with
harmonics $n=2$ (8200\AA), 3 (5600\AA), and 4 (4600\AA).  A period of $\sim$2.6
hr was also derived from a 1.3~mag modulation in the $r$-band light curve.
Because the cyclotron features overwhelm the stellar components in the optical,
this period was ascribed to the rotation of the white dwarf, which was presumed
to be synchronized to the orbital motion. The system was undetected in the RASS
and produced only 0.0012$\pm$0.0003 count s$^{-1}$ in the most sensitive {\it
XMM\/}-Newton detector (the EPIC pn; S04).

A full orbit of circular spectropolarimetry of SDSS~J1324+0320 obtained at the
MMT in 2004 Feb. is presented in Figure 7.  The time sequences represent the
circularly polarized spectral flux\footnote{Note that the polarized flux has
been negated to facilitate comparison with the total flux spectra.}, $-$$v
\times F_\lambda$, and total flux, $F_\lambda$, for successive 13~min
observations. Prior to display, the spectrum of an M6 V star was subtracted
from the data, as this spectral type was determined by S03 to best match
the band features.  It is clear from inspection that the dominant $n=3$
harmonic at 5600\AA\ is indeed the origin of the dramatic modulation in the
$r$-band light curve, but plasma effects cause an even greater variation in
the circularly polarized flux.  In the first and last few observations of the
series an extremely weak, narrow \halpha\ emission line can be recognized in
the total flux spectrum.  The integrated flux and radial velocity of this
feature have been measured for all spectra in which it can be discerned, and
the results are shown in Figure 8. The accompanying least-squares sinusoidal
fits to the variations show that the minimum in \halpha\ flux nearly coincides
with the positive zero-crossing of radial velocity (uncertainties in times of
zero-crossing for the fitted curves are each slightly more than 0.1~hr).  As in
SDSS~J2048+0050, we assign $\varphi=0$ (UT $\approx 10.6$) to this crossing.
Combined with a velocity semi-amplitude of $342\pm146$~km s$^{-1}$, the
characteristics argue for a system viewed at a high inclination, with the line
emission confined largely to the inner hemisphere of the companion.  $\varphi=0$ then corresponds to the time that the secondary star and white dwarf are most
nearly aligned along our line of sight.  The antiphasing between \halpha\ line
strength and overall cyclotron flux in Figure 7 additionally implies that the
dominant accreting pole of SDSS~J1324+0320 is located on the inner hemisphere
of the white dwarf.  As is true for the other low-$\dot m$ systems, the
agreement between the 2.6~hr photometric period and the independently derived
spectroscopic value of $2.62\pm0.24$~hr implies that the white dwarf spin and
binary motion are locked.

The spectral sequences in Figure 7 offer an instructive display of the emission
properties of a magnetized plasma.  The cyclotron absorption coefficient at low
temperatures is such a steep function of frequency in the first few harmonics,
$\alpha\propto{(\omega_c/\omega)}^{10-12}$ (e.g., Meggitt \& Wickramasinghe
1982; Chanmugam et al. 1989), that the observed characteristics can vary
markedly from one harmonic to the next.  As an example, the flux ratio of
harmonic $n=3$ to harmonic 2, $F_3/F_2>1$, signifies moderately high optical
depth in these features, since it would saturate at the Rayleigh-Jeans ratio
$(\omega_3/\omega_2)^3 \approx 3.4$ for $\tau\rightarrow\infty$.  At the same
time, the comparative weakness of the 4600\AA\ harmonic, $F_4/F_3<1$, implies
that the plasma suddenly becomes optically thin at $n=4$.  It is the
combination of the steep harmonic dependence of opacity, low temperature, and
modest column depth that accounts for the cyclotron emission in these systems
being so narrowly confined to just the few lowest harmonics.

A rather high optical depth at $n=2$ is also indicated by the modest
polarization of this harmonic, $\lesssim$10\%, as compared with $v=50-70$\%
reached in harmonic 4. The angular sensitivity of the cyclotron opacity means
that, in the absence of dilution, the smallest optical depth and therefore
largest circular polarization for reasonably thick features like the $n=2$ and
3 features will occur for vantage points that approach the magnetic field
axis.  From the data of Figure 7, this evidently occurs near $\varphi=0$ (10.6
UT). The fact that both the total and polarized flux at $n=4$ are actually
reduced at this phase attests to this feature being optically thin throughout.

Beyond this qualitative understanding, the situation becomes more complex.
Ferrario et al. (2004) encountered difficulties in modeling the detailed
behavior of the harmonics in SDSS~J1553+5516 and SDSS~J1324+0320 when they
noted that the fluxes in harmonics 2 and 3 in Figure 8 are actually {\it
lowest\/} at phases where the optical depth as indicated by the degree of
polarization is {\it largest\/}. This is contrary to predictions of the
standard model. Ferrario et al. obtained some success with a model that
recognizes the effects of ram pressure in compressing the upper white dwarf
atmosphere.  The significance of this fact is that cyclotron emission from the
impact region would then traverse and be attenuated by cooler regions of the
surrounding undisturbed photosphere for a certain range in viewing angle.
Faraday mixing and depolarization might also be important, and for this reason
phase-dependent linear spectropolarimetry of one of these systems would be
valuable.

\subsection{Orbital Variations in SDSS~J1553+5516}

Coarse phase-resolved spectroscopy and a single spectropolarimetric observation
of SDSS~J1553+5516 were presented by S03 and point-source models of
phase-resolved spectropolarimetry were discussed by Ferrario at the 2002
Capetown IAU colloquium (190) on magnetic cataclysmic variables
(unpublished).  Summarizing those conclusions with respect to the new time
series data obtained on 2002 May 8 and shown in Figure 9, cyclotron emission
is arising in a relatively cool plasma, $kT\sim1$~keV, that is moderately
optically thick at the $n=3$ (6200\AA) harmonic, but optically thin at all
phases at $n=4$ (4650\AA). Indeed, defining orbital phase $\varphi=0$ by the
positive zero crossing of emission line radial velocity at 7.9 UT (see below),
the Doppler effect alone would broaden features beyond the narrow structure
that is observed in polarized flux ($v\times F$) near $\varphi=0.75$ and 0.25
(6.5 and 9.0 UT, respectively) unless $kT\lesssim1$~keV. The horn-shaped
appearance of the $n=3$ harmonic in polarized flux evident over a broad
interval around $\varphi=0$ ($7.3-8.6$ UT) is even more prominent in
fractional circular polarization, $v(\%)$. This is characteristic of a plasma
with modest optical depth, where the intensity at the harmonic peak approaches
the blackbody limit in the extraordinary ray (Rousseau et al. 1996; Ferrario
et al. 2004).

Because of its relative brightness, the \halpha\ emission line in
SDSS~J1553+5516 is rather easily measured, and analysis of the data of Figure 9
yields the radial velocity curve depicted in Figure 10.  The phasing and
amplitude ($K=250\pm18$~km~s$^{-1}$) of the line, shown by filled circles, is
shared by the $\lambda$7050 TiO bandhead (crosses), conclusively showing that
both arise on the secondary star.  The fact that the photometric period of 4.39
hr reported by S03 also matches the spectroscopic variations proves the
existence of spin-orbit synchronism.  Finally, we point out that the maximum in
\halpha\ emission-line flux occurs near the end of the sequence in Figure 9
($\varphi\sim0.5$), once again indicating that the chromospheric emission is
concentrated on the inner hemisphere of the companion.

\section{Secondary Stars and Distances}

The secondary stars that are so evident in low-$\dot m$ systems currently
provide our only indicators of distance.  From a comparison of the observed
molecular band spectra to those of main sequence stars, Reimers \& Hagen (1999)
and Reimers et al. (2000) estimated spectral types of M3.5 and M4.5 for the
companions in HS~0922+1333 and WX LMi, respectively, and the observed spectral
fluxes relative to late-type spectral standards with parallax measurements,
lead to distance estimates of $\sim$140 and 190~pc. These values and the quoted
total mass accretion rates are listed in Table 3, though it is useful to point
out the uncertainties in this technique, as Schwarz et al.  (2001) find a
somewhat smaller value ($D\sim100$~pc) for HS~0922+1333.

In applying the same method to the SDSS examples, it appeared that the lack of
readily available, accurately flux-calibrated optical spectra of M dwarf
standards was a significant difficulty, so the calibration observations
described in the Appendix were undertaken.  Those results provide easily
computed absolute monochromatic fluxes in several narrow spectral bands over
the range $4700-7500$\AA\ for M dwarfs from M0 to M6.5. The relations presented
therein were used in the following analysis.

Spectral types for the secondaries of SDSS~J1553+5516 and SDSS~J1324+0320 were
judged by S03 to be M5 and M6, respectively, and we adopt their
classifications.  For SDSS~0837+3830 and SDSS~2048+0050, the spectra in Figure
2 offer good S/N and wide spectral coverage, and comparison of these to the
SDSS M dwarf standards from Hawley et al. (2002) leads to estimates of M5 and
M3, respectively (it is our opinion that uncertainties in the process do not
warrant the claim of fractional subtypes).  Using these spectral types, best
cancellation of the molecular features was found for secondary fractions in
the $7450-7500$\AA\ calibration band of 80\%, 45\%, 60\%, and 90\% for the
stars SDSS~J1553, 1324, 0837, and 2048, respectively.  The resulting spectra
are presented in Figure 11. The use of eq. A1 together with the coefficients in
Table A1 then yield the distances and mass accretion rates listed in Table 3.
For reference, an accretion rate of $10^{-13}~M_\sun$ yr$^{-1}$ corresponds to
a luminosity $L_{acc}\sim6\times10^{29}$ erg s$^{-1}$.  The distance estimate
to SDSS~J1553+5516 is somewhat larger than the 100~pc used by S03, but the
values are probably consistent to within the accuracy of the technique. In
fact, considering the combined uncertainties in spectral classification of the
secondaries, fractional contributions to the spectra, standard calibration, and
slit/fiber losses, distances to individual objects should not be trusted to
better than 50\%, or the resulting luminosities to a factor of two.  Note that
the spectrum of SDSS~J2048+0050 displayed in Figure 11 exhibits a broad bump
around 6600\AA\ that exceeds the residuals from the M-star subtraction and that
we take in support of our polarimetric identification of the $n=3$ cyclotron
harmonic (\S3.2).

In computing the total mass accretion rates, the white dwarf mass was taken to
be 0.6~$M_\sun$ and gravitational energy was assumed to be converted solely
into cyclotron flux (see, however, \S6), with only a small allowance for
harmonics outside the observed window. The flux progression in the observed
harmonics, which typically peaks at $n=3$, supports this view. The exception is
SDSS~0837+3830, which apparently is not optically thin in the bluest harmonic
measured ($n=4$).  For this system the uncertainty in $\dot M$ is somewhat
higher. The absence of significant X-ray emission has been demonstrated for
SDSS~J1553+5516 and SDSS~J1324+0320 (S04), and theory (e.g., Woelk \&
Beuermann 1992) asserts that this should be the case for the field strengths
and specific accretion rates indicated by the cyclotron spectra.

\section{Primary Star Temperatures}

Temperatures for the white dwarfs have been estimated at $T_{\rm eff} \le
10,000$~K for both HS~0922+1333 and SDSS~J1553+5516 (Reimers et al. 1999;
S03), and 13,000~K $\pm$ 1000~K for WX LMi (Reimers \& Hagen 2000).  All
are unusually cool in comparison to the primaries of Polars, which themselves
are cooler than the white dwarfs in nonmagnetic CVs (Sion 1999).  The
difference between the latter two classes is understood to reflect different
levels of accretion-induced heating, largely compressional heating of the
interior (Townsley \& Bildsten 2002).  By extension, the temperature limit for
SDSS~J1553+5516 not only implies that the system is relatively old,
$\tau_{cool} \gtrsim 0.7$~Gyr, but can be used to constrain the mean accretion
rate over the past $\sim$10$^6$~yr to $\langle \dot M \rangle \lesssim
3\times10^{-12}~M_\sun$ yr$^{-1}$ (Townsley \& Bildsten 2004; S04).

Analysis of the spectra of the other three SDSS systems yields similarly low
estimates: $T_{\rm eff} \lesssim 14,000$~K for SDSS~J0837+3830 and
$\lesssim$7,500~K for both SDSS~J1324+0320 and SDSS~J2048+0050.  Each of these
values is based on the observed flux level in a gap between cyclotron
harmonics, after subtraction of the best-fitting M dwarf template, and makes
use of surface fluxes from the nonmagnetic $\log g=8$ DA model atmospheres
summarized by Bergeron et al.  (1995).  We have also assumed a stellar radius
of $8\times10^8$~cm (appropriate for $M_{wd}\sim0.6~M_\sun$), and include an
allowance of 50\% above the distances listed in Table 3 to account for the
uncertainties in those estimates.  While higher mass white dwarfs (smaller
radii) would permit higher temperatures, values increase by only $20-50\%$ for
$M_{wd}=1~M_\sun$, and in any case the possibility of additional emission
sources contaminating the gaps renders the estimates upper limits.

In evaluating the spectra, it was evident that the spectral shape of the
underlying continuum occasionally appeared hotter than the temperature
indicated by the measured flux at a specific wavelength.  The situation is
depicted in Figure 11 as a comparison between the secondary-subtracted SDSS
spectra and model stellar energy distributions.   Because the model
temperatures, which range from 5,500~K for SDSS~J1324+0320 to 9,500~K for
SDSS~J0837+3830, were computed on the basis of the flux at a cyclotron-free
wavelength in the interval $5000-5600$\AA, the continua match here.  However,
for SDSS~J1553+5516 and SDSS~J2048+0050 the model atmospheres fall well short
of the measured flux around 4000\AA.  The spectrum of SDSS~J1324+0320 is too
faint in the blue for a reliable comparison and the broader $n=4$ harmonic in
SDSS~J0837+3830 confounds the assignment of the stellar continuum at the
shortest wavelengths.

The inconsistency between continuum shape and absolute flux level for at least
SDSS~J1553+5516 and SDSS~J2048+0050 is reminiscent of the difficulties that
plague temperature determination of non-accreting highly magnetic white dwarfs,
where observed optical continua are found to be significantly steeper than
predicted by nonmagnetic white dwarf models (e.g., Schmidt et al. 1986;
G\"ansicke et al. 2001).  However, the disparity is considerably greater in the
low-$\dot m$ systems, despite a field strength ($\sim$65~MG) below the level at
which substantial effects on the emitted continuum are expected ($>$100~MG).
Unfortunately, the best attempts to estimate bound-free opacities in the
presence of a strong magnetic field (Merani et al. 1995) have not proven to be
noticeably better in explaining the observed spectral energy and polarization
distributions (Jordan \& Merani 1995) at high strengths.

We suggest that a more viable explanation of the differences between indicators
of stellar temperature is the presence of a heated cap surrounding the magnetic
pole(s).  Irradiated hot spots are known to be important in and around the
impact regions of Polars, where temperatures as high as $3\times10^5$~K and
sizes up to a fraction $f=0.1$ of the stellar surface are measured (e.g.,
Vennes et al. 1995; G\"ansicke et al. 1998; Mauche 1999).  Even if cyclotron
emission is the dominant cooling mechanism in a low-$\dot m$ system,
approximately half of that radiation will be directed downward, where it will
be intercepted and reradiated in the UV or EUV by the white dwarf photosphere.
Considering the sizes of the impact regions on Polars, a hot spot covering
$2-5$\% of the stellar surface in a low-$\dot m$ magnetic binary would not be
unreasonable.  From simple simulations using blackbodies (G\"ansicke et al.
1998; Mauche 1999) we have found that a spot with $T_{\rm eff} = 20,000$~K and
$f=0.04$ could steepen the net continuum of a star with a surface temperature
of 8,500~K to an equivalent temperature of $>$10,000~K, while increasing the
absolute flux in the optical by only $\sim$30\%.  Other combinations are, of
course, possible.

The total luminosity of a hot spot with the above characteristics implies a
total accretion rate $\dot M\sim4\times10^{-13}~M_\sun$ yr$^{-1}$, slightly
higher than the highest rates indicated by the optical cyclotron luminosity in
Table 3.  However, there is independent evidence in favor of such spots in the
form of the flux-modulated narrow Balmer emission lines described in \S3 and
4.  While Balmer emission is a common attribute of active late-type stars, in
each of the four SDSS examples the line flux was found to be enhanced on the
inner (irradiated) side of the secondary.  Assuming that each ionizing photon
incident on the secondary gives rise to one \halpha\ photon, a spot with
$T_{\rm eff}=20,000$, $f=0.04$ could account for a luminosity $L_{\rm H\alpha}
\sim 3\times10^{26}$ erg s$^{-1}$ in a binary with $P=4$~hr.  This is a
significant fraction of the $2-20\times10^{26}$ erg s$^{-1}$ measured in the
modulated portions of the line.  Higher temperatures and lower covering
factors would bring the numbers into better agreement. The fact that
SDSS~J0837+3830 exhibits the largest measured \halpha\ luminosity, as well as
the highest inferred accretion rate and hottest white dwarf of the SDSS
objects, is a reassuring consistency check.  Some support is also provided by
the (marginal) detection of a $\sim$46~eV component in the EUV portion of the
{\it XMM-Newton\/} spectrum of SDSS~J1553+5516 (S04) that could not be
explained by coronal emission from the secondary star.  A more definitive test
is possible in the form of UV/EUVE photometry through an orbital cycle (\`a~la
G\"ansicke et al. 1998), indeed NUV/FUV observations of SDSS~J1553+5516 are in
the queue for {\it GALEX\/}.  If irradiated accretion spots contribute
significantly to the optical continua of the low-$\dot m$ magnetic binaries,
the implication is that the physics of the impact region involve more cooling
mechanisms than have been discussed to date.

\section{Low Accretion-Rate Magnetic Systems as Pre-Polars}

\subsection{Undersized Companions}

The total accretion rate estimates in Table 3, $\sim$$5\times10^{-14} -
3\times10^{-13}~M_\sun$ yr$^{-1}$, are all $<$1\% of the values typically
measured for Polars during high accretion states.  In fact, the numbers bear
more similarity to the integrated solar wind mass loss rate of
$2\times10^{-14}~M_\sun$ yr$^{-1}$.  The tendency of diskless CVs to lapse
into periods of inactivity is well known, but even in low states the accretion
luminosities of Polars are nearly an order of magnitude larger than we have
measured for the low-$\dot m$ systems (Ramsay et al. 2004).  Furthermore, the
durations of Polar low states are typically found to lie in the range of weeks
to $\sim$1 yr\footnote{The glaring exception is the ultrashort period system
EF Eri, which has been in a protracted low state since 1997 (Wheatley \&
Ramsay 1998) but this binary more appropriately belongs near the opposite end
of the evolutionary spectrum from the systems discussed here (Harrison et al.
2004).}, and the overall duty-cycle of accretion appears to be $\sim$50\%
(Hessman et al.  2000; Ramsay et al. 2004).  In contrast, none of the objects
listed in Table 3 has ever been observed in a high state, despite repeated
telescopic visits separated by intervals as long as a few yr (Table 1 and
Schwarz et al. 2001).  These facts, coupled with the low surface temperatures
found for the primary stars, motivate their identity as a new class of
chronically low-$\dot m$ systems.

An explanation for the distinctively low accretion rates and a clue to their
evolutionary status is the fact that in all six cases the secondary stars
appear to underfill their Roche lobes.  This has already been noted for
WX LMi by Schwarz et al. (2001), who pointed out that a star with the
measured spectral type of M4.5 V has a radius 20\% smaller than the Roche lobe
for $P=2.8$~hr.  The lobe dimensions scale directly with stellar separation,
therefore the overall situation can be appreciated simply by noting that the
spectral type expected for a Roche lobe-filling main-sequence star varies from
M2 at a period of 4~hr to M4.5 at $P=2$~hr.  These types are systematically
earlier (by 0.5 to 3 subtypes) than the measured spectral types in Table 3.
Moreover, the mean difference of nearly 1.5 subtypes is large enough to
accommodate any modern mass-radius relation for the lower main sequence as
well as primary star masses as large as the Chandrasekhar limit. Beuermann et
al. (1998) and Baraffe \& Kolb (2000) have pointed out that, among CVs in
general, a spectral type difference in the same sense is observed for systems
with $P>3$~hr. Simulations by Baraffe \& Kolb indicate that the general trend
observed for $3<P<6$~hr, as well as the width of the $2-3$~hr period gap, can
be explained by the donor stars being out of thermal equilibrium owing to mass
loss at a rate $\dot M = 1-2\times10^{-9}$~$M_\sun$~yr$^{-1}$.  While this
accretion rate is in rough agreement with what is inferred from luminosity
estimates for disk CVs above the gap, the generally much lower luminosities of
Polars have long been interpreted to imply that they accrete at rates $1-2$
orders of magnitude less, a fact explicable by the inhibiting effects of the
primary star's magnetic field on magnetic braking from the donor (Webbink \&
Wickramasinghe 2002 and references therein).  Thus, in the Baraffe \& Kolb
(2000) scenario, the systematic difference in spectral type should not exist
for secondary stars in the synchronized magnetic CVs. Unfortunately, Polars are
strongly clustered below the period gap, and only 4 systems exist with $P>3$~hr
in the list analyzed in the above studies.  Among them is V1309 Ori at
$P=8$~hr, whose M0.5 secondary is evolved (Garnavich et al. 1994). There is a
hint that companions in the remaining 3 magnetic objects are $0.5-1$ subtype
later than expected for main sequence donors, but a larger sample is required
to confirm or reject the notion.

Questions about the secondaries in Polars notwithstanding, mass loss rates of
$<$10$^{-12}~M_\sun$ yr$^{-1}$ imply evolutionary timescales far longer than
the donor star's thermal timescale.  For the same accretion rates, irradiation
of the secondary star amounts to at most a few percent of its normal
photospheric output, even for the shortest orbital periods.  Thus, there is
every reason to assume that the secondaries in low-$\dot m$ systems have the
properties of main-sequence stars.

\subsection{A ``Magnetic Siphon'' of the Stellar Wind}

The diminutive sizes of the companions in low-$\dot m$ systems coupled with
evidence that the binaries have been in protracted states of very weak
accretion strongly suggest that some or all of the systems have never achieved
Roche-lobe contact - i.e., that they are {\it pre\/}-Polars.  The absence of
detached magnetic white dwarf + M star binaries has been a subject of recent
discussion (Silvestri et al. 2005; Liebert et al. 2005), with the realization
that selection effects must be playing a significant role.  Identifying the
low-$\dot m$ magnetic binaries as the immediate precursors of at least some
Polars would solve part of the riddle, leaving open the question of the
whereabouts of long-period systems that never experience the common-envelope
phase of evolution.  In this picture, LARPs, while most definitely
``Polar''ized, are not CVs at all, but pre-CVs.

In a binary that has not yet evolved to the point of Roche-lobe overflow, some
portion of the secondary star's stellar wind will always arrive at the surface
of the white dwarf\footnote{This is the same stellar wind that is held
accountable for the dominant angular momentum loss in long-period disk CVs.}.
For nonmagnetic primaries, the accreted fraction is very small, and the
gravitational energy is an insignificant heating source spread over a large
fraction of the white dwarf surface.  However, a strongly magnetic primary
couples effectively to the magnetic field lines of the secondary (Li et al.
1994), so plasma that would otherwise be centrifugally driven to large
distances ends up on the white dwarf.  Moreover, Li et al. find that, above a
critical magnetic field strength, this ``magnetic siphon''\footnote{An apropos
term coined by R. Webbink.} is capable of collecting the {\it entire\/} stellar
wind from the secondary.   Webbink \& Wickramasinghe (2004) offer an analytic
argument based on energy densities that arrives at the same conclusion.  The
critical field strength depends on orbital period and geometry (Li et al.
1995), but appears to be in the range $50-100$~MG. Thus, at $B\sim60$~MG, it is
quite possible that the low-$\dot m$ systems discussed here represent examples
where the magnetic siphon is essentially perfect. Ironically, the effects of
the magnetic field are essential both in capturing the wind and in confining
the resulting emission to a few very prominent features in the optical. If
either process were not active, the objects would not stand out in optical AGN
surveys, and thus would not be isolated for study.  Selection effects in the
SDSS are discussed at more length below.

It is important to note that, while stellar chromospheric/coronal activity and
rotational spin-down are intensely studied effects of mass loss from late-type
stars, thus far only upper limits ($\lesssim$$10^{-10}~M_\sun$ yr$^{-1}$) have
been established for the actual wind rates (e.g., Wargelin \& Drake 2001).  The
mass accretion rates inferred for low-$\dot m$ magnetic accretion binaries may
therefore prove to be the first true measures of the amount of material carried
away.  To apply the results to single stars, corrections will have to be made
for the facts that the binary examples present a reduced gravitational barrier
and are in forced rapid rotation, but even these adjustments will be
unnecessary for evaluating recent suggestions that the long-held magnetic
braking model for CV evolution may be much less effective than previously
thought (cf. Pinsonneault et al. 2002; Kolb 2002).

\subsection{Thoughts on Evolution}

The low-$\dot m$ systems offer interesting insight into the possible
evolutionary stages of a magnetic binary.  We make the standard assumption that
a pre-CV emerges from the common envelope (CE) detached and with the
degenerate core in asynchronous rotation.  Whether the orbital period of a
post-CE binary is affected by the presence of a magnetic field on the
degenerate core is a matter of some debate (Liebert et al. 2005). However,
assuming that tidal forces on the secondary are sufficiently strong, the
binary orbit will decay and the period decrease under the relatively vigorous
effects of magnetic braking via the secondary star's entrained wind (e.g.,
Verbunt \& Zwaan 1981).  The next major event - synchronization of the white
dwarf spin and orbital periods, or the initiation of accretion by Roche-lobe
overflow - depends on the primary star magnetic field strength and secondary
size:  a modest magnetic field paired with a comparatively massive secondary
will initiate Roche-lobe overflow first and appear at relatively long periods
as an Intermediate Polar or other disk CV not yet recognized to be magnetic.
In this case, a magnetic interaction between the stars may eventually
synchronize the system at a shorter period and a Polar will ensue. If,
however, the field strength on the white dwarf is high and/or the secondary
star small, the locked status of the pre-Polars tells us that synchronization
can occur {\it prior\/} to the onset of Roche-lobe overflow, at a period of at
least 4.4~hr for $B=60$~MG.  As the white dwarf spin and orbital periods
approach a common value, and while the stellar wind can still couple to open
field lines of the white dwarf, spin-down might be assisted by the cooperative
magnetic braking mechanism originally suggested by Schmidt et al. (1986) but
shown to be ineffective once a synchronized state is achieved (Li et al. 1994).

It is difficult to imagine a magnetic siphon operating efficiently in a binary
whose spin and orbital motions are uncoupled.  Therefore, we take the onset of
synchronism to be accompanied by the disappearance of angular momentum loss by
magnetic braking.  The estimate of Webbink \& Wickramasinghe (2004) that this
might occur in binaries as large as $a \sim 10~R_\sun$ is probably optimistic,
but it suggests the potential effectiveness of the process.  The pre-Polar now
enters a possibly protracted era of period evolution governed solely by
gravitational radiation and accreting by a magnetic siphon effect on the
stellar wind.  Interestingly, nova eruptions may actually occur during this
phase, but the recurrence time would be exceedingly long ($\sim$1 Gyr).
Eventually, Roche-lobe contact is established in this already synchronized
binary, and a Polar is formed.  Our estimate of $T_{\rm eff}\le5,500$~K of
SDSS~J1324+0320 at $P=2.6$~hr demonstrates that the total time elapsed between
the common envelope and Polar states can exceed 4 Gyr. This, together with the
existence of an Intermediate Polar stage for other systems, offers a ready
explanation for the propensity of Polars to exist at short orbital periods.

\section{Discovery and Census of Pre-Polars}

The identification of 4 pre-Polars in the first $\sim$5000 deg$^2$ surveyed by
the SDSS suggests that an additional $1-2$ will be found before completion of
the project and that $\sim$30 would be cataloged if the entire sky were
surveyed.  From Table 3, the effective sampling distance of such a survey might
be estimated at $D\sim300$~pc, i.e., similar to the volume surveyed for X-ray
bright (Polar) systems by {\it ROSAT\/} (Beuermann \& Burwitz 1995), which
discovered $\sim$80\% of the 80 known Polars.  Thus, at face value, the space
density of systems in the pre-Polar state might be expected to be about half of
that for currently accreting Polars.  The ratio is similar to that implied by a
comparison between 4 pre-Polars and the 14 confirmed Polars that have been
found in the portion of the sky sampled through SDSS DR4 (Szkody et al. 2005
and references therein).

However, questions of completeness must be raised.  First, from the discussion
in \S7, the detectability of pre-Polars is expected to decline rapidly with
increasing orbital period.  If magnetic energy density is the key criterion
(e.g., Webbink \& Wickramasinghe 2004), the $r^{-3}$ decline in field strength
of a dipole implies that the effectiveness of the magnetic siphon will scale as
$\sim$$P^{-4}$.  At the other extreme, the probability of the secondary in a
pre-Polar contacting its Roche surface increases as the period decreases,
meaning that only pre-Polars with tiny secondaries (like SDSS~J1324+0320) will
be found at very short periods, and those will be produced only after long
episodes of slow evolution.  The magnetic field on the white dwarf is also an
important variable, since the weaker-field systems will tend to reach Roche
contact prior to synchronizing, and an Intermediate Polar {\it vs.\/} pre-Polar
will characterize the phase leading up to the synchronized state.

Second, selection effects within the surveys can be important.  The SDSS is
aimed at being complete only for QSOs and galaxies (Stoughton et al. 2002).
One fiber per spectroscopic plug-plate is dedicated to CVs and in practice
samples objects in the white dwarf + pair region of the color-color planes.
Another category for standards is a source of hot white dwarfs.  Nevertheless,
objects that fall in the QSO locus have the highest probability of being
targeted because of the much larger number of fibers available.  In order to
estimate the effects of these constraints on the completeness of a putative
population of pre-Polars, we have computed the SDSS colors of a synthetic
spectrum consisting of a white dwarf, M star, and cyclotron emission for field
strengths ranging from $0-200$~MG.  The measured spectra of SDSS~J1553+5516
and SDSS~J1324+0320 were used to form a cyclotron template with realistic
harmonic intensities and intensity ratios, but the spacing and width (in \AA)
of the harmonics were varied according to the assumed field.  Broadband colors
in the SDSS system were then computed and run through the (complex) SDSS QSO
targeting algorithm (Richards et al. 2002) to determine which regimes in field
strength would be most likely to result in objects selected for fiber
spectroscopy and lead to recognition as pre-Polars.

As an example, we have overplotted the results of our simulation in the $u-g$,
$g-r$ plane in Figure 1, with points separated by equal intervals in $\log B$
and fiducial field strengths (in MG) indicated by numbers adjacent to the
track.  Remember that there are 3 other color-color planes involved in the
targeting decisions not shown here.  At comparatively weak magnetic fields, all
of the low-$n$ cyclotron harmonics lie in the IR and the object is rejected
from spectroscopic targeting by the QSO algorithm as the white dwarf + M star
pair that it is.  At $B=30$~MG, the $n=4$ harmonic enters the $z$ photometric
band and the object is briefly targeted as a high$-z$ QSO.  However, from our
previous discussion on the sensitivity to field strength, such a system might
not achieve a synchronized/siphoned state as a detached binary.  As $B$
continues to increase, the object approaches the stellar locus and loses
targeting until the QSO algorithm is once again triggered for virtually the
entire interval $50<B<90$~MG.  This is the region in which all four SDSS
pre-Polars have been found.  For $90<B\lesssim200$~MG, brief intervals exist
where the binary again receives high- or low-$z$ QSO targeting, but for the
most part it is rejected due to the proximity of the track to white dwarfs and
other excluded classes of stars.  Above $B\sim400$~MG (should such systems
exist), the cyclotron fundamental shifts shortward of the $u$ photometric band,
and the object returns to obscurity as a white dwarf + M pair.  Note that the
strongest magnetic field yet discovered on a Polar is 230~MG on AR UMa (Schmidt
et al. 1996).

Because SDSS~J1324+0320 and SDSS~J1553+5516 were selected to form the spectral
template, it is no accident that these two systems lie close to the simulation
path at approximately the correct field strength.  However, the wide variety of
spectra displayed by the examples makes it impossible to accurately model the
entire picture.  For example, the dominant $n=4$ harmonic in SDSS~J0837+3830
places it solidly within the low-$z$ QSO domain of the $u-g$, $g-r$ plane, and
it was targeted as such.  The single weak harmonic around near 4500\AA\ in
SDSS~J2048+0050 is not sufficient to move it far from the white dwarf + M star
pairs.  It was not targeted by the QSO algorithm, but rather as an F
turnoff/subdwarf star candidate in the SEGUE extension aimed at studying the
Milky Way.

It therefore appears possible that various physical and observational biases
could preferentially select SDSS pre-Polar systems with magnetic field
strengths in a broad interval around 60~MG.  The fact that the field strength
at the dominant pole of each of the six binaries discovered thus far, including
the two from the Hamburg Schmidt survey, falls within just $\pm5\%$ of a common
value might be an artifact of small number statistics.  However, we would
expect to discover systems with $B\sim50$ or 80~MG soon.  If selection effects
are as important as they appear to be, the total population of pre-Polars could
be much larger than was postulated at the beginning of this section.
Conceivably, this could strain estimates of the space density of magnetic CVs
as derived from the census of Polars, but careful attention must be paid to the
several Gyr that an object can hide as a detached magnetic binary with an orbit
very slowly decaying by gravitational radiation.

\acknowledgements{The authors are grateful to P. Smith for assistance at the
telescope and to X. Fan and G. Richards for running the synthetic colors
through the SDSS QSO targeting algorithm.  Valuable discussions were enjoyed
with L. Ferrario, J. Liebert, and D. Wickramasinghe, and R. Webbink offered
key insights as well as a careful reading of the manuscript. Funding for
the creation and distribution of the SDSS Archive has been provided by the
Alfred P. Sloan Foundation, the Participating Institutions, the National
Aeronautics and Space Administration, the National Science Foundation, the
U.S.  Department of Energy, the Japanese Monbukagakusho, and the Max Planck
Society.  The SDSS Web site is http://www.sdss.org/.  The SDSS is managed by
the Astrophysical Research Consortium (ARC) for the Participating Institutions.
The Participating Institutions are The University of Chicago, Fermilab, the
Institute for Advanced Study, the Japan Participation Group, The Johns Hopkins
University, the Korean Scientist Group, Los Alamos National Laboratory, the
Max-Planck-Institute for Astronomy (MPIA), the Max-Planck-Institute for
Astrophysics (MPA), New Mexico State University, University of Pittsburgh,
University of Portsmouth, Princeton University, the United States Naval
Observatory, and the University of Washington.  Support is provided by the NSF
for the study of magnetic stars and stellar systems at the University of
Arizona through grant AST 03-06080, and for cataclysmic variables at the
University of Washington through AST 02-05875.}


\clearpage

\begin{deluxetable}{lcclcl}

\tablecaption{LOG OF OBSERVATIONS}

\tablewidth{6in}

\tablehead{\colhead{Object} &
\colhead{UT Date} &
\colhead{Telescope} &
\colhead{Type} &
\colhead{Duration} &
\colhead{Comments} \\
\colhead{(SDSS+)} &
\colhead{(yyyymmdd)} &
\colhead{} &
\colhead{} &
\colhead{(h:mm)} &
\colhead{} }

\startdata
J083751.00+383012.5 & 20020212 & SDSS 2.5~m & Spectroscopy     & 0:45 & \\
                    & 20040424 & MMT        & Cir. spectropol. & 1:13 & \\
                    & 20040513 & MMT        & Cir. spectropol. & 1:27 & \\
                    & 20040921 & MMT        & Cir. spectropol. & 0:35 & \\
                    & 20041130 & MDM 2.4~m  & CCD imaging      & 2:56 & poor seeing \\
                    & 20041201 & MDM 2.4~m  & CCD imaging      & 4:44 & poor seeing \\
                    & 20050316 & Bok 2.3~m  & Cir. spectropol. & 0:30 & \\
\\
J132411.57+032050.5 & 20030529 & MMT        & Cir. spectropol. & 2:28 & \\
                    & 20040216 & MMT        & Cir. spectropol. & 2:59 & \\
\\
J155331.12+551614.5 & 20020318 & Bok 2.3~m  & Cir. spectropol. & 3:27 & clouds at end \\
                    & 20020508 & Bok 2.3~m  & Cir. spectropol. & 4:16 & \\
                    & 20040216 & MMT        & Cir. spectropol. & 0:18 & \\
\\
J204827.91+005008.9 & 20040824 & SDSS 2.5~m & Spectroscopy     & 1:33 & \\
                    & 20040916 & Bok 2.3~m  & Cir. spectropol. & 1:12 & cirrus \\
                    & 20041004 & APO 3.5~m  & Spectroscopy     & 4:18 & \\
\enddata




\end{deluxetable}

\begin{deluxetable}{lccccc}

\tablecaption{SDSS PHOTOMETRY}

\tablewidth{5in}

\tablehead{\colhead{Object} &
\colhead{$g$} &
\colhead{$u-g$} &
\colhead{$g-r$} &
\colhead{$r-i$} &
\colhead{$i-z$} }

\startdata
SDSS J083751.00+383012.5 & 19.14 & +0.01 & +0.03 & +0.49 & +0.70 \\
SDSS J204827.91+005008.9 & 19.38 & +0.56 & +0.70 & +1.08 & +0.68 \\
\enddata




\end{deluxetable}

\begin{deluxetable}{lcccrccl}

\tablecaption{LOW ACCRETION-RATE MAGNETIC BINARIES}

\tablewidth{6.in}
\tablehead{\colhead{Object} &
\colhead{Period} &
\colhead{Mag.\tablenotemark{a}} &
\colhead{$B$\tablenotemark{b}} &
\colhead{$\dot M$} &
\colhead{$D$} &
\colhead{Sp. Type} &
\colhead{Ref.} \\
\colhead{ } &
\colhead{(h)} &
\colhead{ } &
\colhead{(MG)} &
\colhead{($M_\sun$~yr$^{-1}$)} &
\colhead{(pc)} &
\colhead{(secondary)}
}

\startdata
SDSS~J1324+0320 & 2.60         & 22.1 & 64      & $\sim$$1\times10^{-13}$ & 450 & M6~~\, & 5,6,7 \\
WX LMi (HS~1023+3900)          & 2.78         & 18.0 & 60(68)  & $<$$3\times10^{-13}$    & 140 & M4.5   & 1,3,4 \\
SDSS~J0837+3830 & 3.18 or 3.65 & 19.1 & 65(??)  & $\sim$$2\times10^{-13}$ & 330 & M5~~\, & 7 \\
HS~0922+1333             & 4.07         & 19~~ & 66(81:) & $\sim$$3\times10^{-13}$ & 190 & M3.5   & 2,4 \\
SDSS~J2048+0050 & 4.2~\,       & 19.7 & 62:     & $\sim$$5\times10^{-14}$ & 260 & M3~~\, & 7 \\
SDSS~J1553+5516 & 4.39         & 18.5 & 60      &       $6\times10^{-14}$ & 130 & M5~~\, & 5,6,7 \\
\enddata


\tablenotetext{a}{$g$ for SDSS objects; $B$ for HS objects.}
\tablenotetext{b}{Parentheses denote secondary pole.}
\tablerefs{
(1) Reimers et al. (1999);
(2) Reimers \& Hagen (2000);
(3) Schwarz et al. (2001);
(4) Schwope et al. (2002);
(5) Szkody et al. (2003);
(6) Szkody et al. (2004);
(7) This paper. }

\end{deluxetable}

\clearpage

\begin{figure}
\includegraphics{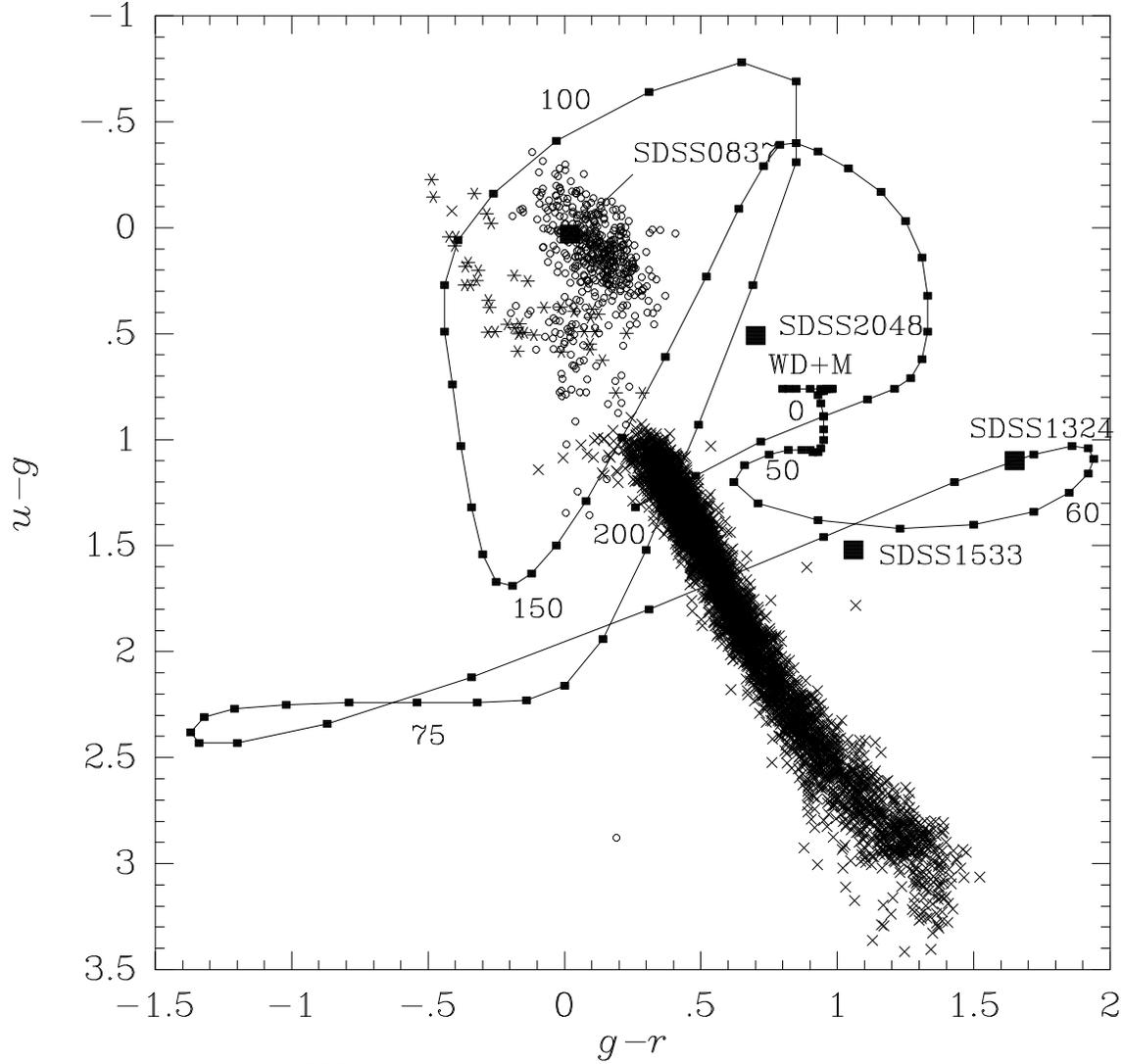}
\vspace{4.5truein}

\figcaption{Locations of the four SDSS low-$\dot m$ magnetic accretion
binaries in the $u-g$, $g-r$ color-color plane.  Also shown are disk stars
{\em (crosses)}, white dwarfs {\em (stars)}, and QSOs {\em (circles)}, taken
from the North Galactic Pole SDSS color simulations of Fan (1999).  The
approximate location of white dwarf + M star pairs is indicated, but the QSO
targeting algorithm actually defines this excluded region according to $g-r$,
$r-i$ colors. Only objects with $g<21$ are shown, and measured color bands
differ slightly from those used in the Fan simulations, but the data suffice
for illustrative purposes.  The serpentine path depicts the track of a model
low-$\dot m$ magnetic system with field strength varying from $0-200$~MG, as
marked and discussed in \S8.}
\end{figure}

\clearpage

\begin{figure}
\includegraphics{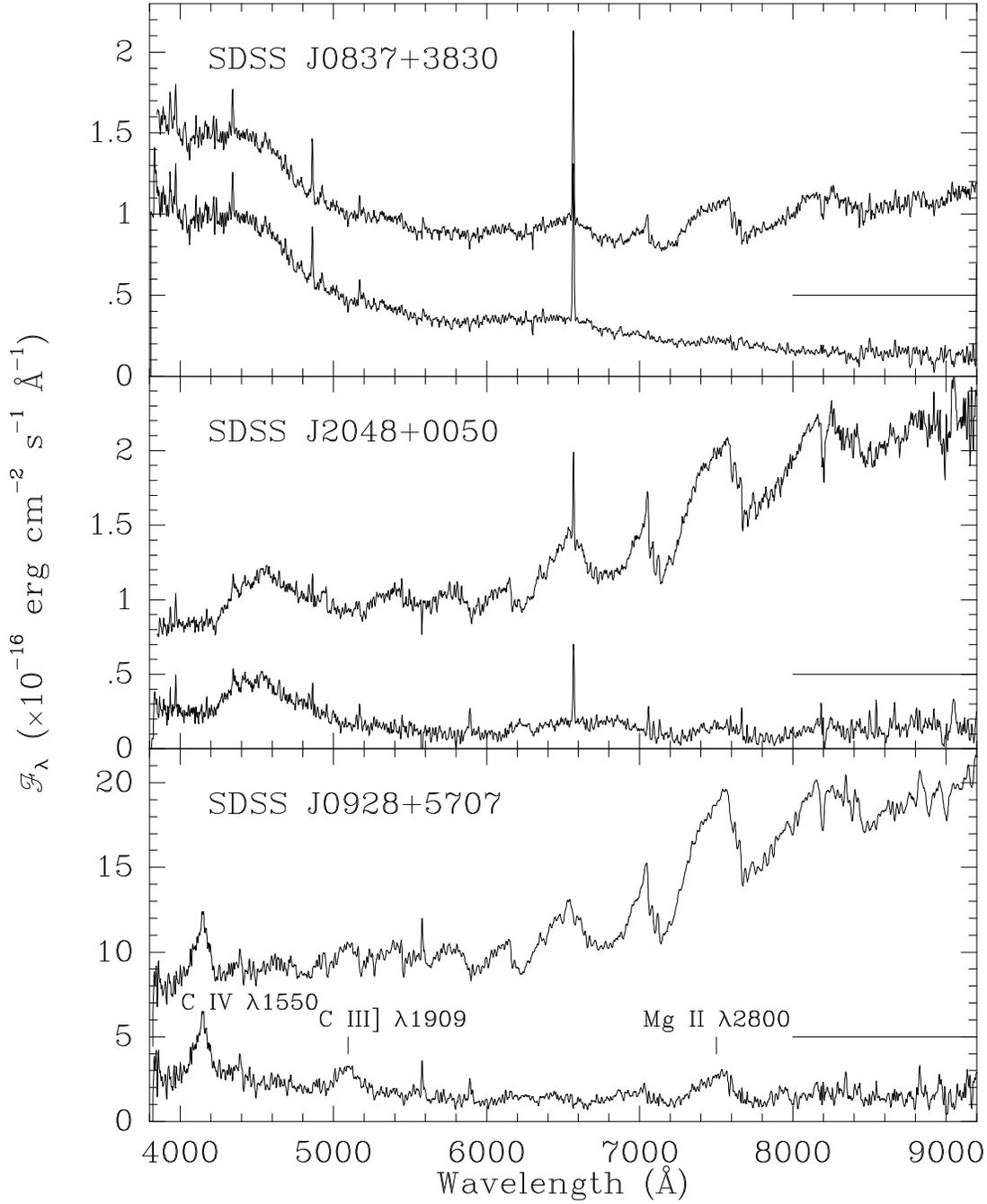}
\vspace{5.5truein}

\figcaption{{\it (Top and middle, bold)\/}: Observed SDSS spectra of the two
new magnetic accretion binaries. Each exhibits a clear hump in the blue that is
found to be strongly polarized, plus an M-star continuum at longer
wavelengths.  Spectra are displaced upward for clarity, with the zero-point of
the flux scale indicated at right.  {\it (Narrow)\/}: Spectra after
subtraction of the best-fitting late-type main-sequence star.  See text for
details. {\it (Bottom, bold)\/}: The spectrum of SDSS~J0928+5707 appears
qualitatively similar to those in the upper panels, but it is unpolarized. {\it
(Narrow)\/}: Subtraction of an M3 V spectrum reveals that the object is a
superposition with a QSO at $z=1.67$.  Comparison of the observed spectra of
SDSS~J0837+3830 and SDSS~J2048+0050 with the examples from Szkody et al.
(2003) attest to the wide range in spectral properties displayed by the class.}
\end{figure}

\clearpage

\begin{figure}
\includegraphics{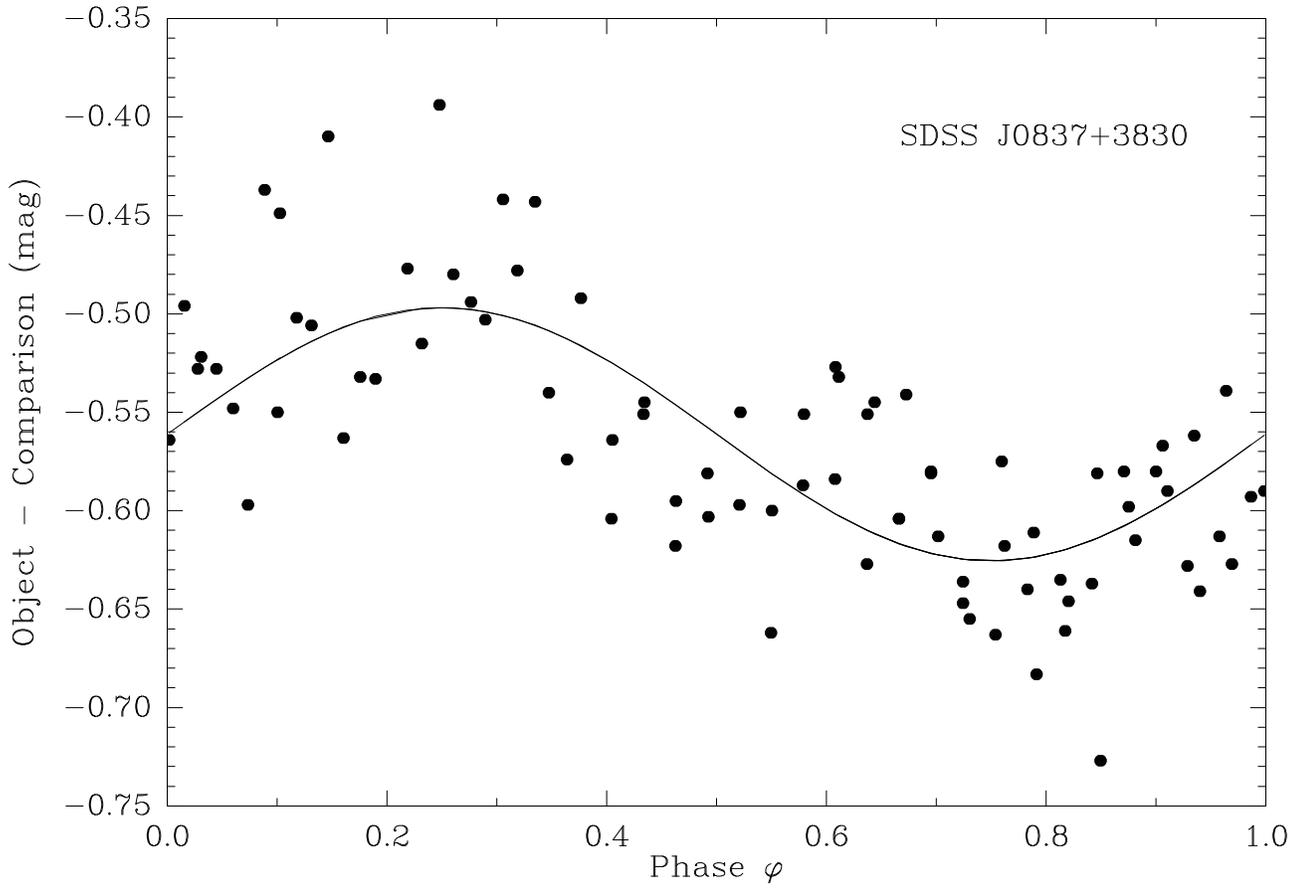}
\vspace{2.25truein}

\figcaption{Two consecutive nights of CCD photometry of SDSS~J0837+3830 phased
on a period of 3.18~hr.  The least-squares fit sine wave semiamplitude is 0.06
mag.  Scatter is attributable to poor seeing during the observations.}

\end{figure}

\clearpage

\begin{figure}
\includegraphics{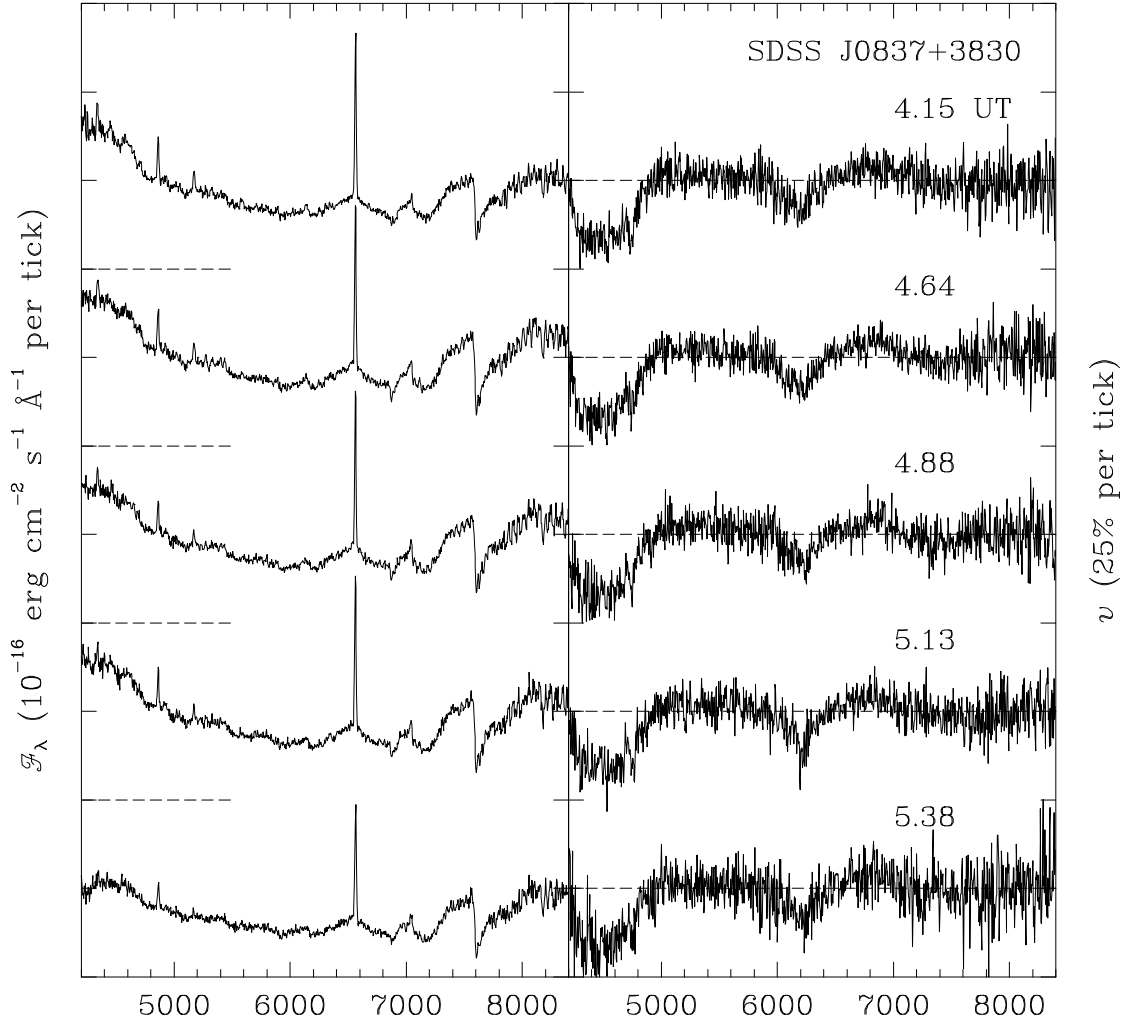}
\vspace{3.85truein}

\figcaption{Spectral flux {\it (left)\/} and circular polarization {\it
(right)\/} for SDSS~J0837+3830 on 2004 May 13.  Sequences cover 1.23 hr, or
nearly 40\% of the orbital period, with the mid-UT of each spectrum indicated
in the right panel.  The only convincing change through the series is a
decrease in the emission-line strength, likely due to our varying view of the
inner hemisphere of the secondary star.  Note the distorted appearance in
polarization of the harmonic at 6200\AA\ and the weak positive circular
polarization between harmonics, indicating that accretion occurs onto both
magnetic poles.}
\end{figure}

\clearpage

\begin{figure}
\includegraphics{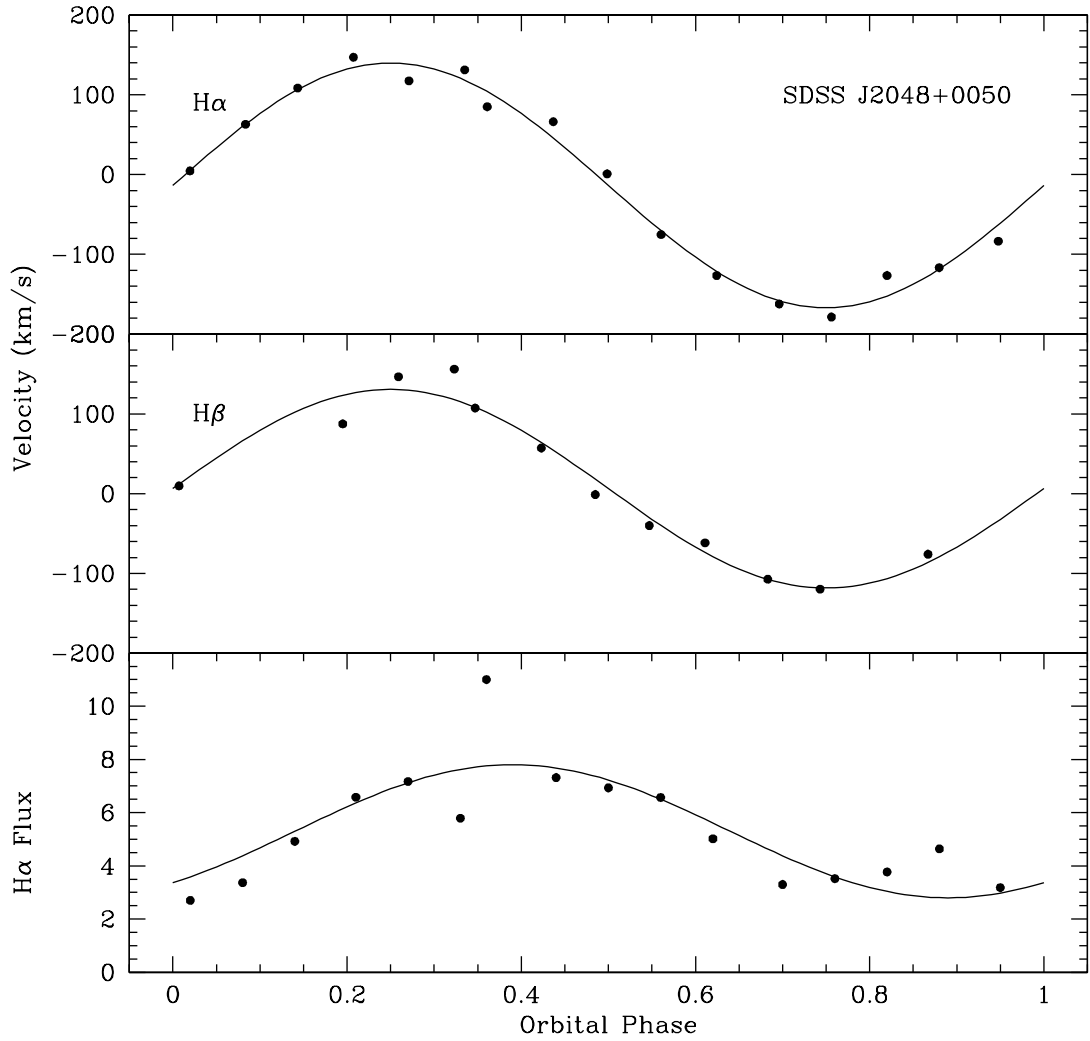}
\vspace{3.5truein}

\figcaption{Emission-line velocities and \halpha\ flux (in units of 10$^{-16}$
erg cm$^{-2}$ s$^{-1}$) for SDSS~J2048+0050, phased on a period of 4.2 hr, with
$\varphi=0$ defined by the positive zero-crossing of radial velocity.}

\end{figure}

\clearpage

\begin{figure}
\includegraphics{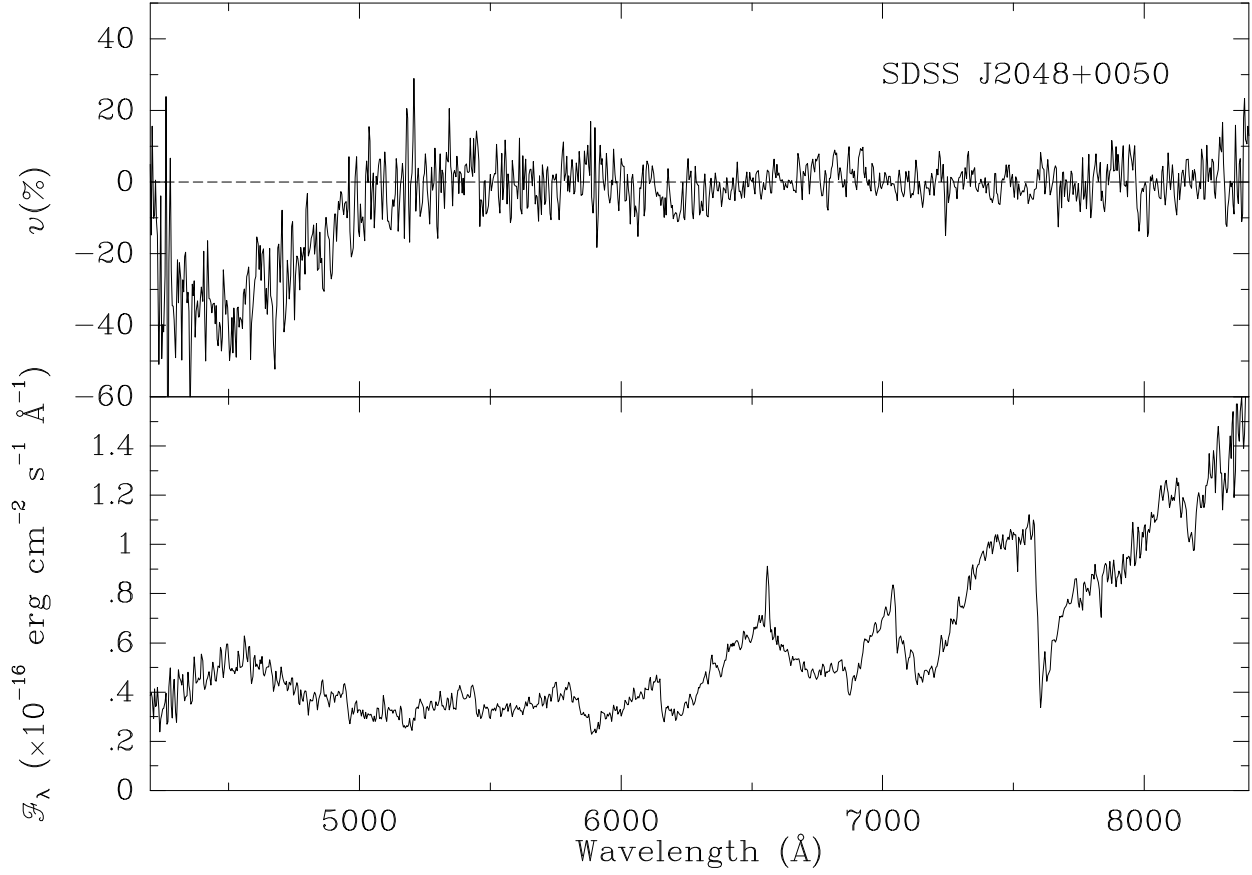}
\vspace{3.truein}

\figcaption{Spectral flux {\it (bottom)\/} and circular polarization {\it
(top)\/} for SDSS~J2048+0050 on 2004 Sep. 16.  In addition to the
strongly-polarized cyclotron hump around $\lambda$4550, a very weak harmonic
may be present near 6200\AA. This would indicate a magnetic field strength of
$\sim$62~MG. The data represent slightly more than 1 hr of exposure, or about
one-quarter of the period. The measured polarization in the 4550\AA\ harmonic,
coupled with the strength of the hump in total flux, implies that the
cyclotron emission in this feature is essentially 100\% circularly polarized.}

\end{figure}

\clearpage

\begin{figure}
\includegraphics{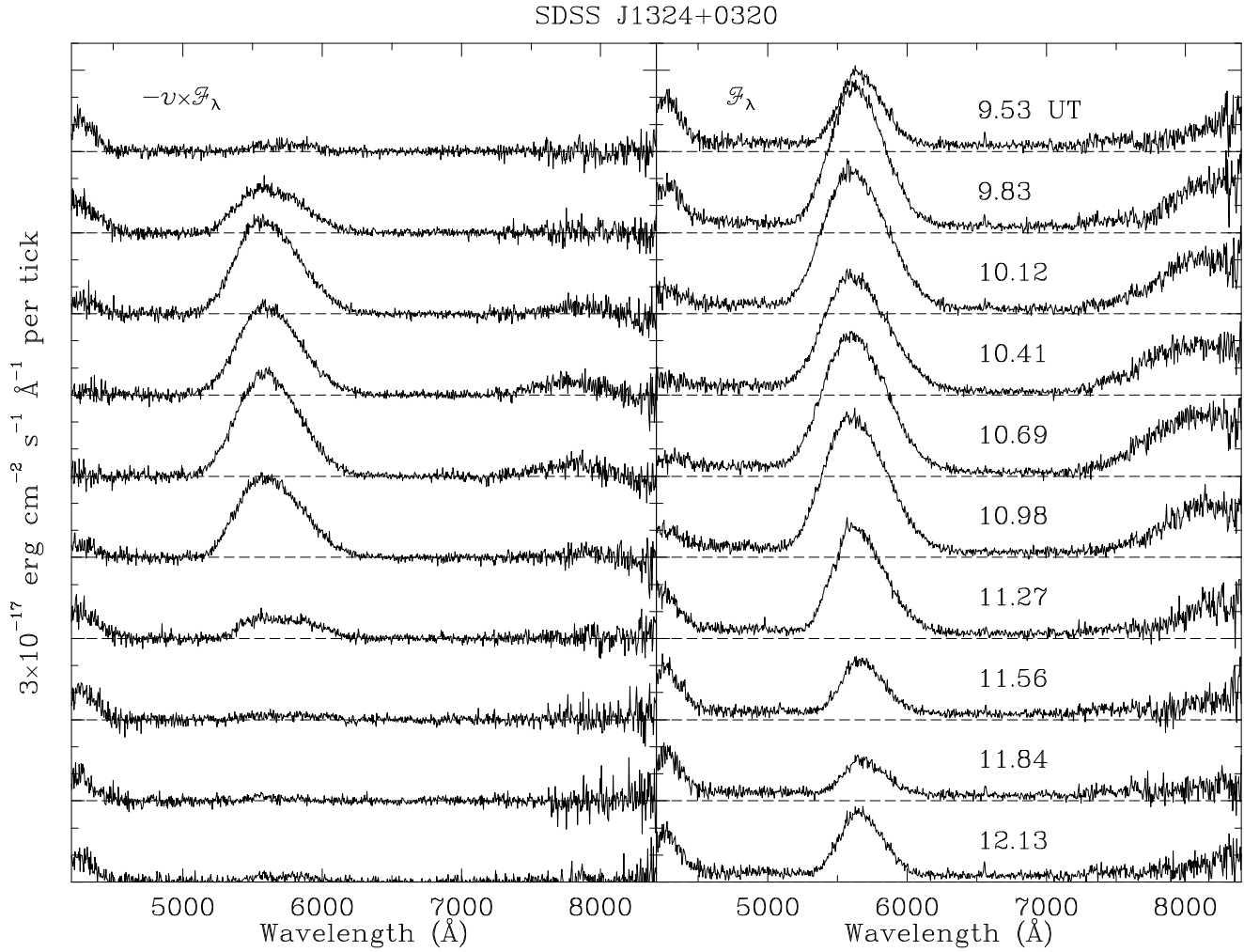}
\vspace{3.25truein}

\figcaption{Spectrum of circularly polarized flux, $-$$v\times F_\lambda$ {\it
(left)\/}, and total flux $F_\lambda$ {\it (right)\/}, {\it vs.\/} UT as
marked, covering a full 2.6 hr cycle of SDSS~J1324+0320 on 2004 Feb. 16.  The
total flux spectrum is shown after the subtraction of an M6 V stellar
template, the spectral type judged to best match the observed stellar
features.  Note that the polarized flux has been negated to facilitate
comparison with the total flux spectrum.}

\end{figure}

\clearpage

\begin{figure}
\includegraphics{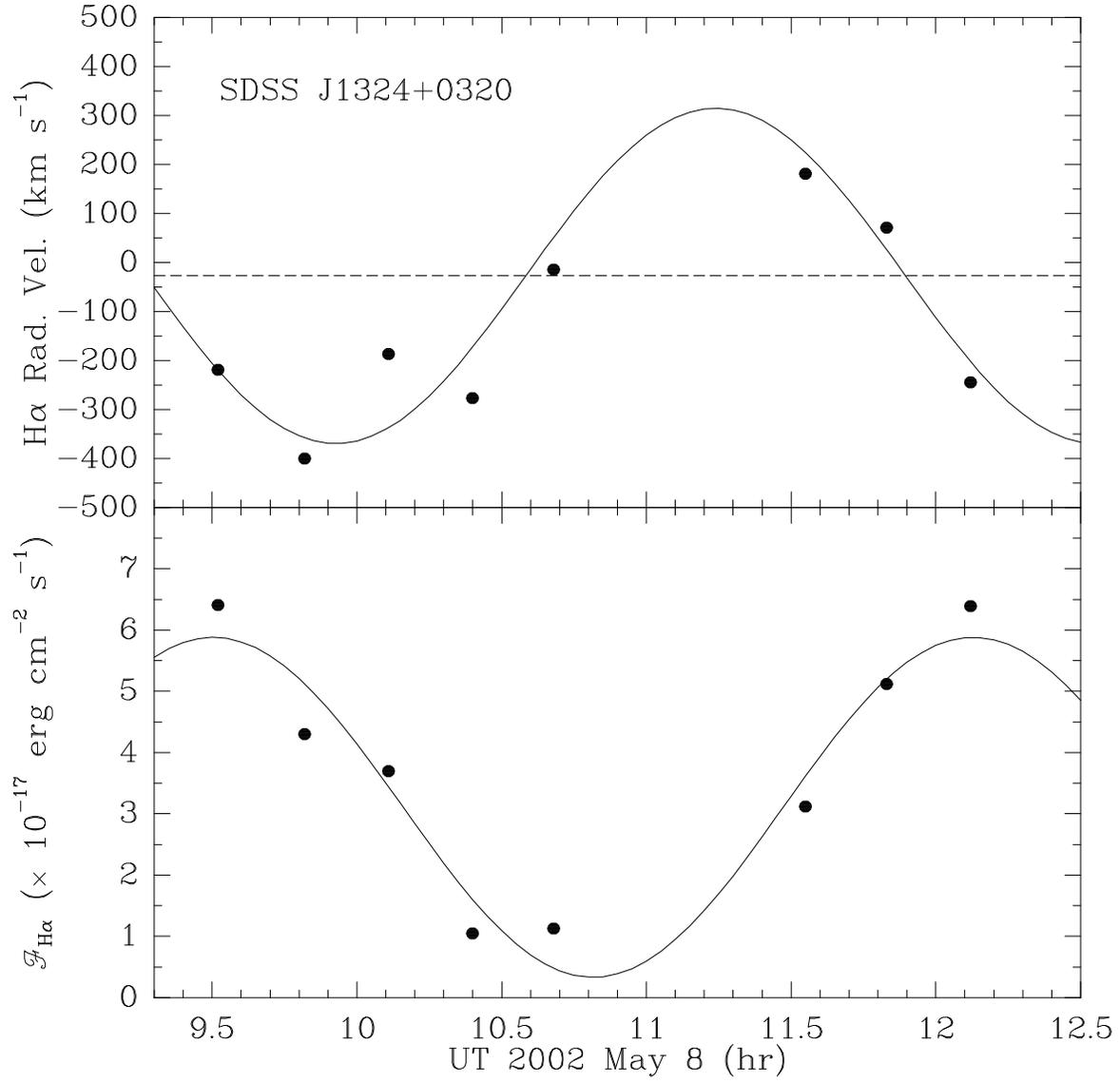}
\vspace{5.truein}

\figcaption{Radial velocity {\it (top)\/} and flux {\it (bottom)\/} of the
weak, narrow \halpha\ emission line of SDSS~J1324+0320 {\it vs.\/} UT
from the data of Figure 7.  The best-fit period (shown) has a value of
$2.62\pm0.24$ hr.  Note that the minimum \halpha\ flux and positive
zero-crossing of radial velocity coincide with the peak in flux of the $n=3$
cyclotron harmonic in Figure 6, indicating that the spin and orbital periods
are essentially locked and that the emission line arises on the inner
hemisphere of the secondary star.}

\end{figure}

\clearpage

\begin{figure}
\includegraphics{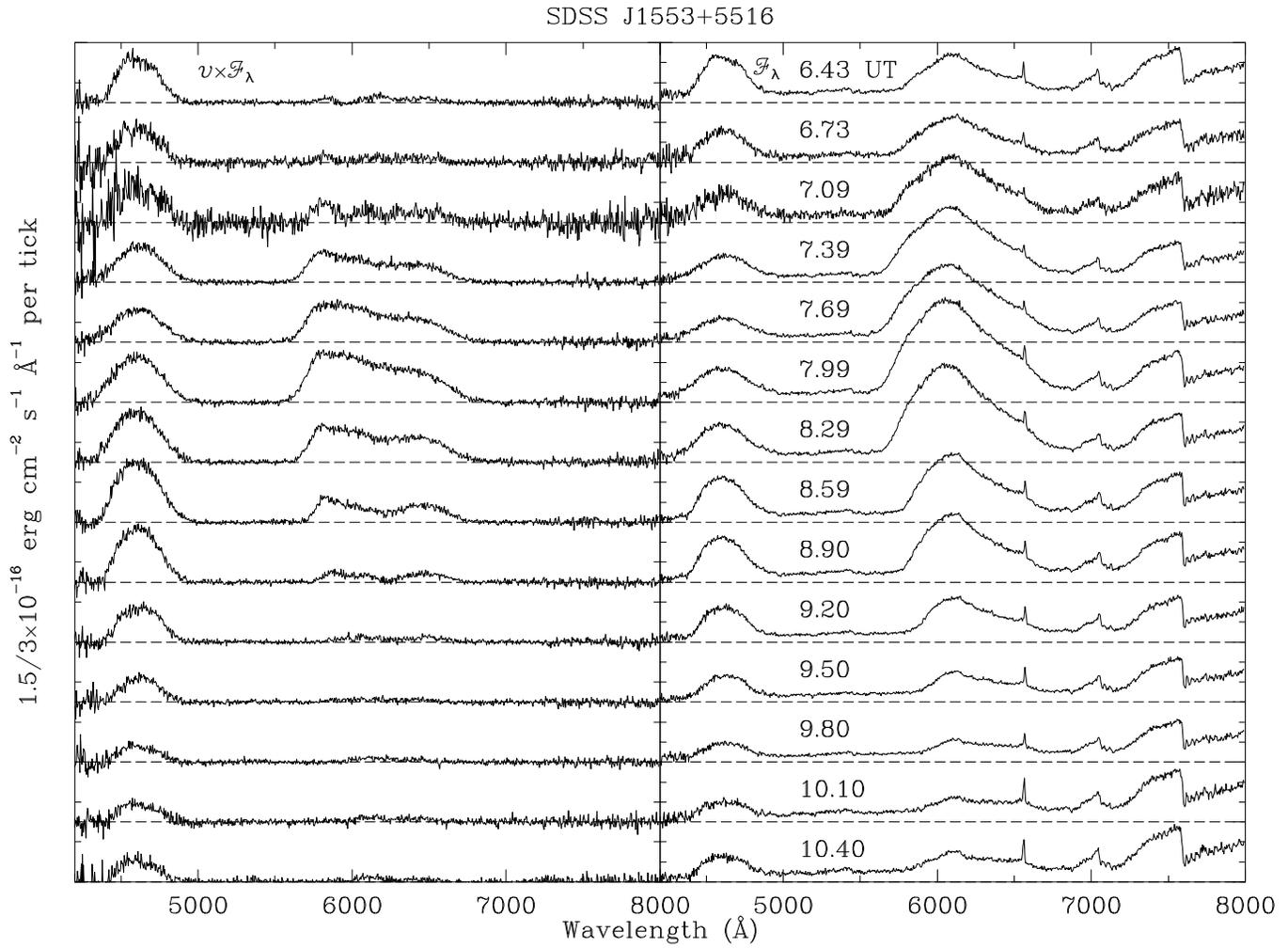}
\vspace{3.25truein}

\figcaption{Spectrum of circularly polarized flux, $v\times F_\lambda$ {\it
(left)\/}, and total flux $F_\lambda$ {\it (right)\/}, vs. UT as marked,
covering nearly a complete cycle of SDSS~J1553+5516 on 2002 May 8. Scaling of
the ordinate differs by a factor 2 between the two panels, as indicated.}

\end{figure}

\clearpage

\begin{figure}
\includegraphics{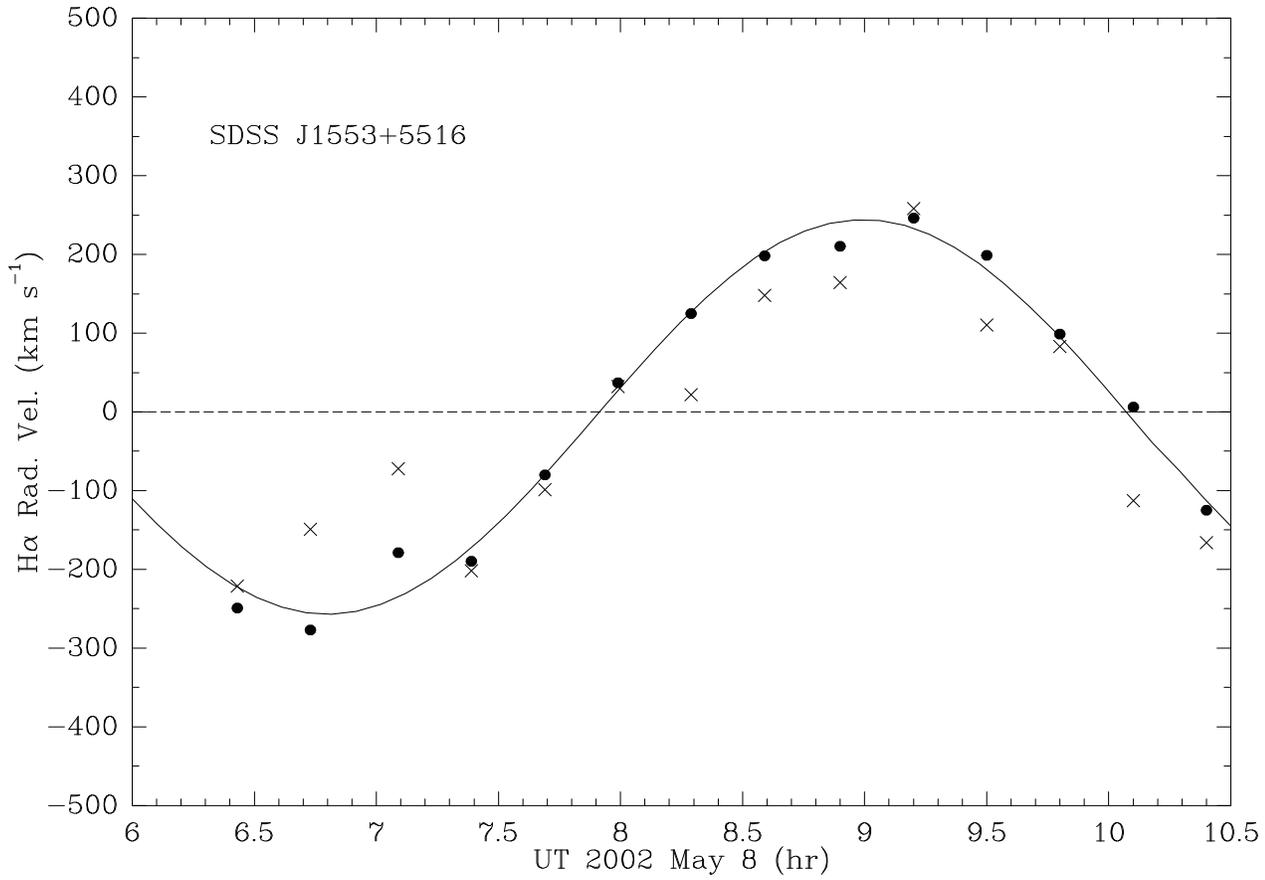}
\vspace{3.truein}

\figcaption{Radial velocity of the \halpha\ emission line {\it (circles)\/}
and 7050\AA\ TiO bandhead {\it (crosses)\/} of SDSS~J1553+5516 {\it vs.\/} UT
from the data of Figure 9.  The common phasing and semiamplitude (250~km
s$^{-1}$) of the two features confirm that the emission lines originate on
the secondary star.}

\end{figure}

\clearpage

\begin{figure}
\includegraphics{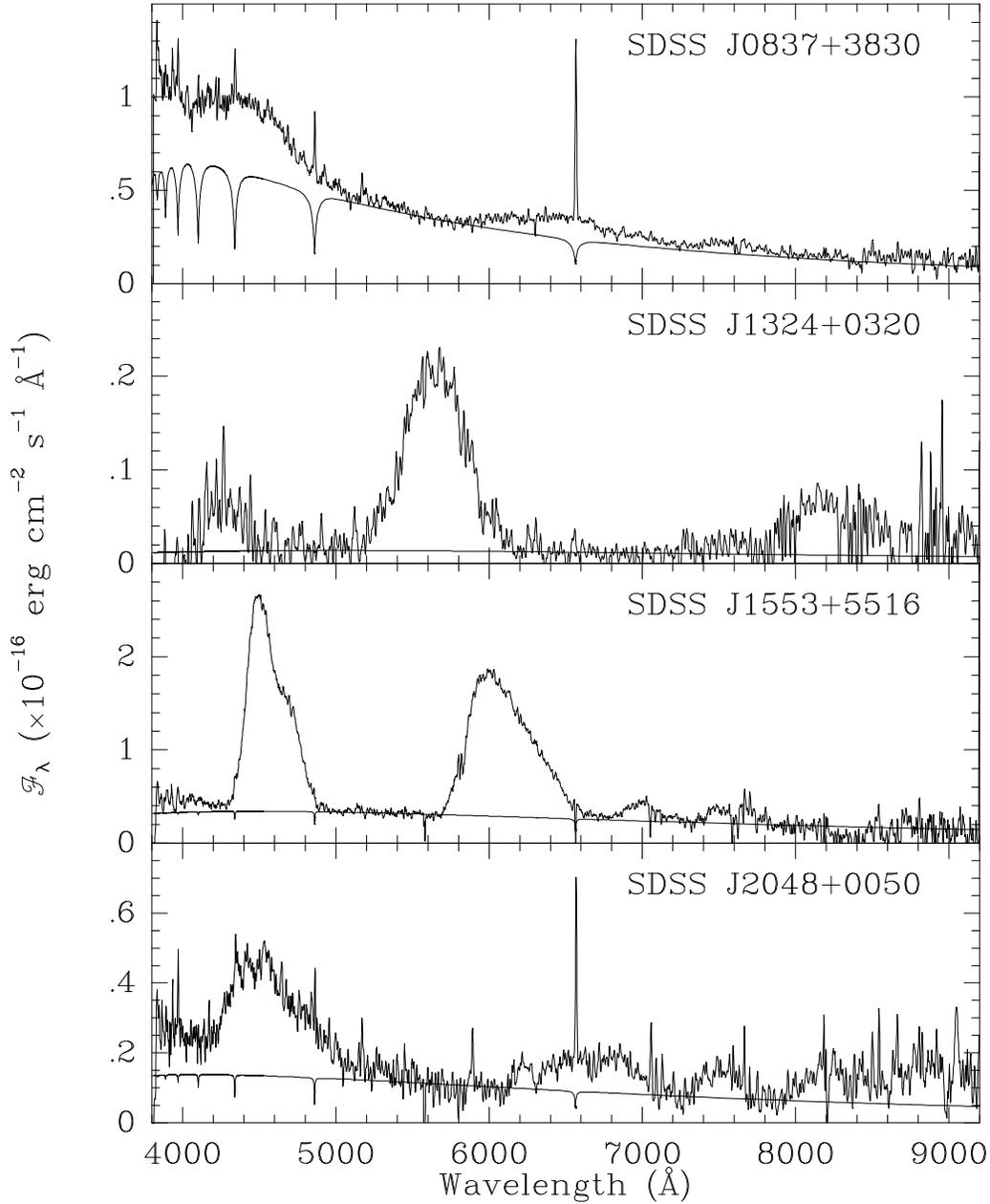}
\vspace{6truein}

\figcaption{Secondary-subtracted survey spectra of the four SDSS low-$\dot m$
systems {\it (bold)\/}, with the energy distributions of white dwarf
photospheres plotted below {\it (narrow)\/}, where the stellar temperatures
have been computed according to the measured spectral flux between cyclotron
harmonics in the $5000-5600$\AA\ interval.  The implied white dwarfs are
unusually cool for accretion binaries, with $T_{\rm eff}$ ranging from 5,500~K
for SDSS|J1324+0320 to 9,500~K for SDSS~J0837+3830.  Note, however, that the
stellar continuum falls short of the observed flux near 4000\AA\ in at least
SDSS~J1553+5516 and SDSS~J2048+0050, suggesting the presence of an irradiated
hot spot surrounding the pole(s).}

\end{figure}

\clearpage

\appendix

\section{Distance Estimation from New Calibrated M Dwarf Spectra}

While the specific accretion rate (e.g., gm cm$^{-2}$ s$^{-1}$) in a magnetic
binary can be determined from spectroscopic analysis and/or an X-ray to
optical flux ratio, a total mass-transfer rate requires distance information.
Often, the only clue to this comes from the spectrum of a stellar component
coupled with an assumption as to the structural state of the star.
Disregarding indications to the contrary at certain orbital periods (Beuermann
et al. 1998 and references therein), the ZAMS relation is often assumed for
secondaries in CVs, and in the absence of strong radiative heating, spectral
types and luminosities appropriate to main-sequence stars are also taken to
apply.  In the case of chronically very low accretion-rate magnetic binaries,
where mass transfer does not occur through Roche-lobe overflow and there is
good reason to believe that contact with the Roche surface has never been
established, mass loss has not had the opportunity to affect stellar
structure, and the assumption of main-sequence properties are especially
secure. Unfortunately, even though spectral types of late-type dwarfs are
generally based on optical line and band ratios (Kirkpatrick et al. 1991),
absolute energy outputs of the stars are generally given in broad-band optical
and/or infrared absolute magnitudes (e.g., Henry et al. 1994; Baraffe et al.
1998; Hawley et al. 2002).  The use of these to estimate distances to new
objects solely from optical spectra not only involves computational steps, but
the process can be problematic for binaries where additional emission
components are present (primary star, accretion disk, cyclotron radiation).
Moreover, available spectral atlases (e.g., Gunn \& Stryker 1983; Jacoby et
al. 1984) often do not extend sufficiently far into the red to enable
comparison, or were taken with rather narrow entrance slits or under
non-photometric conditions, so absolute flux calibrations cannot be trusted.
It therefore seemed appropriate to acquire a new sequence of M dwarf spectra
for the express purpose of establishing monochromatic flux trends suitable for
distance estimation to short-period binaries.

Fourteen stars with spectral types M0 to M6.5 V were chosen for observation
from the RECONS\footnote{Research Consortium on Nearby Stars;
http://www.chara.gsu.edu/RECONS.} list of the 100 nearest stars.  The stars
observed are: G099$-$049, GJ~205, GJ~229A, GJ~273, GJ~285, GJ~299, GJ~338A,
GJ~406, GJ~408, GJ~411, GJ~412A, GJ~412B, GJ~1111, and GJ~1116A.  Distances are
known from parallax measurements to an uncertainty of 5\% for all cases, and to
$<$2\% in the vast majority.  The entire list was observed in a 2 hr period of
steady, photometric conditions with 1\farcs5 FWHM seeing with the 2.3~m Bok
telescope using the spectropolarimeter operating as a simple spectrograph. The
region $4200-8200$\AA\ was available in a single grating setting at a
resolution of $\sim$15\AA.  An entrance aperture of width 5\arcsec\ width
ensured low slit losses, and absolute spectral response functions were based
on observations of 3 standard stars calibrated by Massey et al. (1988),
measured both before and after the program stars.  Atmospheric extinction
corrections used mean coefficients for Kitt Peak, but in the interests of
efficiency the terrestrial O$_2$ band features in the red were not removed.
This would have required the observation of an additional star with a
featureless spectrum near each of the program targets.

Spectra that sample the range in spectral type are shown in Figure A1.  To
facilitate their use in estimating distances, five narrow bands were selected
in regions where the flux is not changing strongly with wavelength and away
from deep molecular bands, chromospheric emission lines, and terrestrial
absorption features.  For the band redward of 7000\AA, an additional
requirement was that the bandwidth be broader than the fringing period caused
by optical interference in the CCD.  Mean spectral fluxes (in units of erg
cm$^{-2}$ s$^{-1}$ \AA$^{-1}$) in these bands were computed for each star and
the values adjusted to a common distance of 10~pc.  The results are displayed
in Figure A2 for all 14 stars, using spectral types provided on the RECONS web
page.  Also shown are least-squares fits to relations of the form
\begin{equation}
\log(F_\lambda) = A + B \cdot S^C
\end{equation}
where $S$ is the M subtype.  For all bands, it was found that a value of
$C=1.6$ adequately reflects the curvature in the relations, and least-squares
values for the fitting coefficients $A$ and $B$ are given in Table A1.

The mean dispersion around the fits amounts to slightly less than 50\% in flux,
equivalent to a distance uncertainty of somewhat more than 20\%.  This is
considerably greater than the distance uncertainty in any single star, and is
far too large to be explained by flux calibration errors in the calibration
spectra.  Instead, the dispersion is likely attributable to differences in
spectral classification (typically equivalent to an uncertainty of $\sim$0.5
subtype), plus possibly real variations among the stars within a subtype, and
reflects the limitations of the technique.

\clearpage

\begin{deluxetable}{lcc}
\tablenum{A1}
\tablecaption{FITTING COEFFICIENTS FOR MONOCHROMATIC FLUX RELATIONS}

\tablewidth{2.truein}
\setlength{\tabcolsep}{0.08in}

\tablehead{\colhead{Band} &
\colhead{$A$\tablenotemark{a}} &
\colhead{$B$\tablenotemark{a}}
}

\startdata
$4700-4740$ & $-$12.23 & $-$0.179 \\
$5100-5150$ & $-$12.29 & $-$0.165 \\
$6100-6140$ & $-$11.97 & $-$0.174 \\
$6760-6840$ & $-$12.04 & $-$0.159 \\
$7450-7500$ & $-$11.85 & $-$0.122 \\
\enddata


\tablenotetext{a}{$F_\lambda$ in units of erg cm$^{-2}$ s$^{-1}$ \AA$^{-1}$,
for a distance of 10~pc.}

\end{deluxetable}


\vskip .5truein

\begin{figure}
\figurenum{A1}
\includegraphics{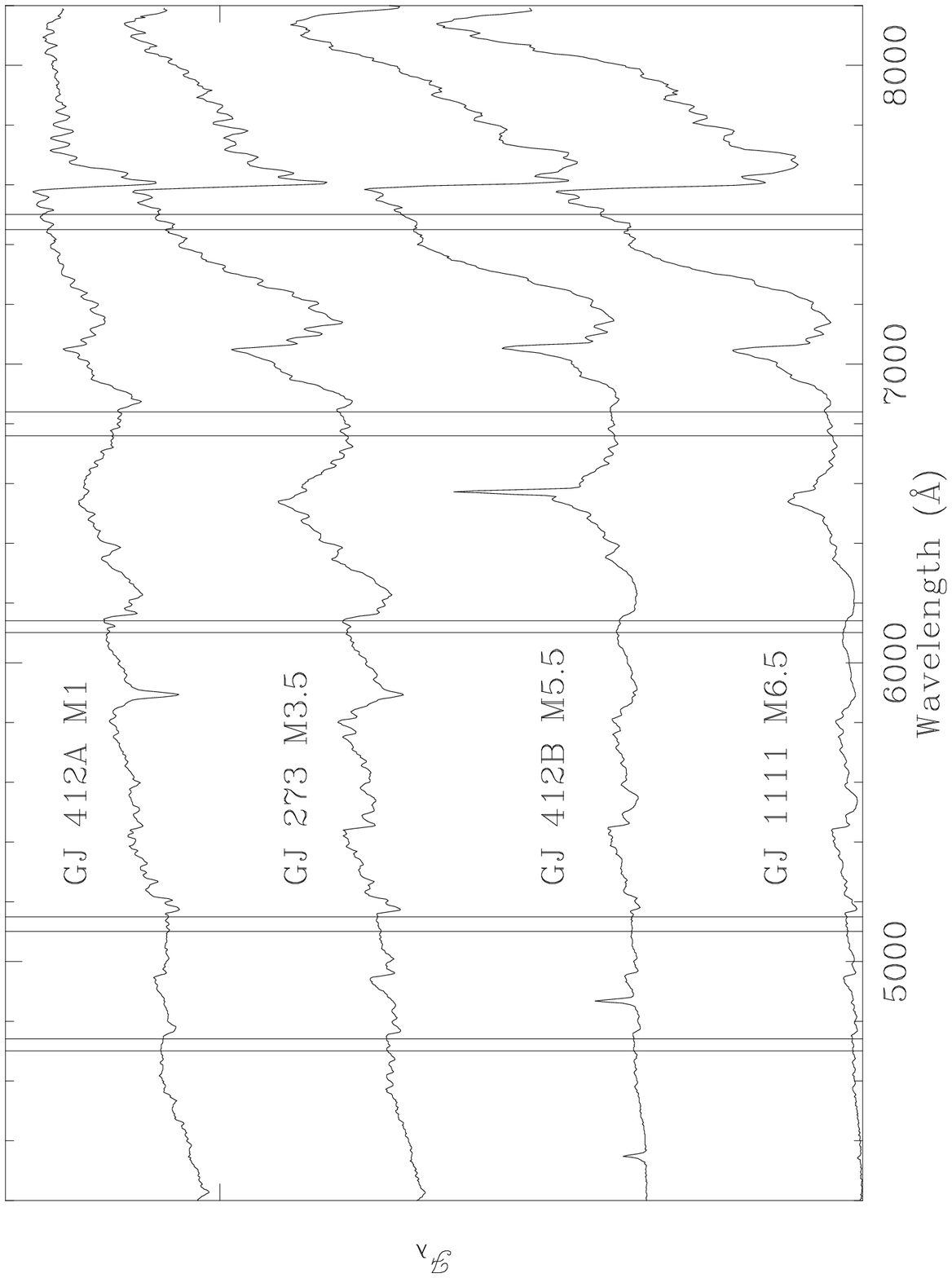}
\vspace{2.truein}

\figcaption{Sample M dwarf spectra obtained with a resolution of $\sim$15\AA\
to calibrate monochromatic absolute flux-spectral type relations.  Narrow
bands defined for the analysis are marked.}

\end{figure}

\clearpage

\begin{figure}
\figurenum{A2}
\includegraphics{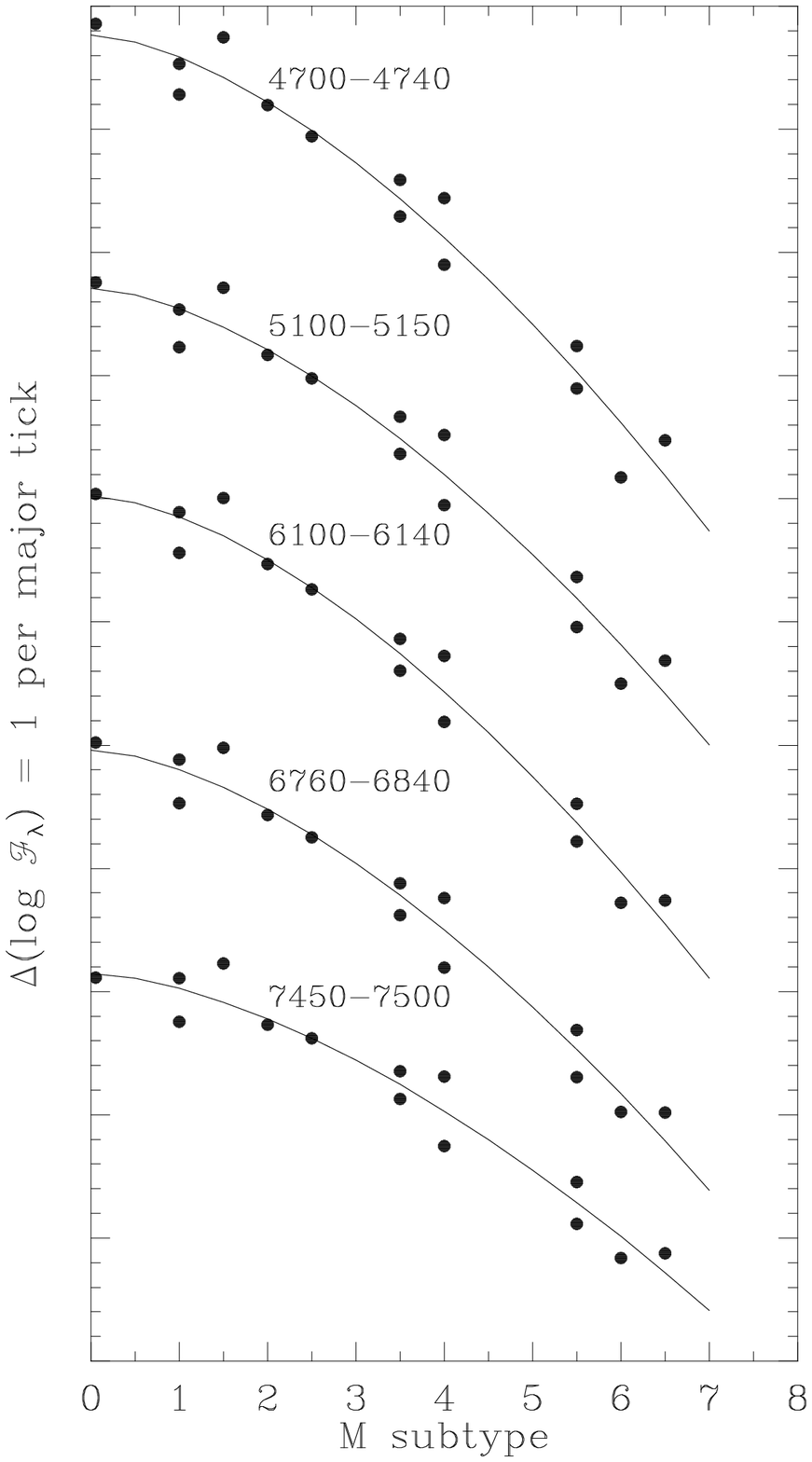}
\vspace{6.truein}

\figcaption{Dependence of spectral flux in the narrow bands of Figure A1 -
adjusted to a common distance of 10~pc - on M subtype, with a displacement of
2 dex between successive relations.  Equations describing the fits are provided
in the text.}

\end{figure}

\end{document}